\documentclass[aps, prd, twocolumn, superscriptaddress]{revtex4-1}
\usepackage{graphicx}
\usepackage{amssymb}
\usepackage{amsmath}
\usepackage{booktabs}
\usepackage[colorlinks=True,linkcolor=blue,anchorcolor=blue,citecolor=blue,CJKbookmarks=true]{hyperref}
\usepackage{bm}

\usepackage{ulem}

\begin{document}
\title{Nonadiabatic quantum Vlasov equation in spinor QED}

\author{Z. L. Li}
\email{zlli@cumtb.edu.cn}
\affiliation{School of Science, China University of Mining and Technology, Beijing 100083, China}
\affiliation{State Key Laboratory for GeoMechanics and Deep Underground Engineering, China University of Mining and Technology, Beijing 100083, China}

\author{Y. J. Li}
\email{lyj@aphy.iphy.ac.cn}
\affiliation{School of Science, China University of Mining and Technology, Beijing 100083, China}
\affiliation{State Key Laboratory for GeoMechanics and Deep Underground Engineering, China University of Mining and Technology, Beijing 100083, China}

\date{\today}

\begin{abstract}
The nonadiabatic quantum Vlasov equation in spinor QED is derived, and its relation to the well-known adiabatic one is established by three methods. One is by an explicitly analytical expression, the second is by the Dirac equation in the V gauge, and the last is by introducing a turn-off electric field. Wherein what the first two of them are given is an instantaneous relation. Moreover, the time evolution of the distribution function for a specific momentum and the momentum distribution of created particle pairs after turning off the electric field are calculated and compared with those in scalar QED. It is found that both the oscillation periods of the distribution functions in spinor and scalar QED equal pi divided by the total energy of a particle after the electric field is turned off. The momentum distributions in spinor and scalar QED show a novel oscillation and out-of-phase behavior that cannot be explained by the Stokes phenomenon. These findings will further deepen our understanding of the quantum Vlasov equation and its application in vacuum pair production.
\end{abstract}

\maketitle

\section{INTRODUCTION}

The understanding of vacuum began a long time ago, but it was not until the establishment of quantum mechanics that people gained a deeper understanding of it. Due to the seminal works of Dirac \cite{Dirac19301}, Sauter \cite{Sauter1931}, Heisenberg and Euler \cite{Heisenberg1936}, and Schwinger\cite{Schwinger1951}, vacuum decaying into the electron-positron pairs in the presence of a strong electric field has become one of the most important theoretical predictions of quantum electrodynamics (QED). With the rapid development of laser technology, especially the chirped pulse amplification (CPA), the electric field strength of the laser pulse is approaching the critical one, $E_{\rm{cr}}=m^2/e\approx 1.3\times10^{18}\rm{V}/\rm{m}$ (where $m$ and $e$ are the mass and charge magnitude of the electron, respectively. Note that the natural units $\hbar=c=1$ are used throughout this paper), in which the vacuum can produce observable electron-positron pairs in experiments. Therefore, vacuum pair production has attracted a lot of research interest \cite{Xie2017, Fedotov2023}.

The commonly used methods of vacuum pair production are S-matrix \cite{Brezin1970,Hu2010,Ilderton2011,Piazza2016,Mackenroth2018}, Wenzel-Kramers-Brillouin-like (WKB-like) approximation \cite{Marinov1977,Dumlu2010,Li2014-1,Oertel2019,Taya2021,Kohlfurst2022}, worldline instanton technique \cite{Affleck1982,Kim2002,Dunne2005,Dumlu2011,Schneider2018}, quantum Vlasov equation (QVE) \cite{Kluger1998,Schmidt1998,Alkofer2001,HebenstreitPRL2009,Hebenstreit2009,Kohlfurst2014}, Dirac-Heisenberg-Wigner (DHW) formalism \cite{Bialynicki-Birula1991,Hebenstreit2010,Hebenstreit2011,Blinne2014,Li2017,Kohlfurst2020}, and computational quantum field theory (CQFT) \cite{Cheng2010,Jiang2012,Su2012,Dong2017,Su2019}. Wherein the QVE is a very powerful tool to study vacuum pair production for a spatially homogeneous and time-dependent electric field with linear polarization. By using it, the non-Markovian effect \cite{Schmidt1999}, back reactions and damping \cite{Bloch1999,Jiang2023}, dynamically assisted Schwinger pair production \cite{Orthaber2011,Nuriman2012}, optimization and enhancement of pair production \cite{Kohlfurst2013,Hebenstreit2014,Li2014}, pair production in chirped electric fields \cite{Dumlu20101,Jiang2013,Abdukerim2017,GongPRD2020}, and so on are studied. Its equivalence with the quantum mechanical scattering approach \cite{Dumlu2009}, DHW formalism \cite{Hebenstreit2010,Li2019}, and CQFT \cite{Li2021} are also proved.

The QVE used in the above studies is called adiabatic QVE. Another version of QVE put forward in \cite{Kim2011} is called nonadiabatic QVE (NAQVE). In \cite{Huet2014}, it was found that if the frequency of the initial state is not equal to that of the final one, the asymptotic distribution function for NAQVE will oscillate with time and not converge, but for QVE it is constant. This difference reflects the distinction between the in-in and in-out formalisms because NAQVE corresponds to the in-in formalism while QVE is related to the in-out formalism.
Interestingly, the in-in formalism is widely used to study cosmological problems \cite{Campos1994,Higuchi2011,Ota2023}. That is because, in cosmology, the asymptotic out particle state is absent in general. Sometimes, even the asymptotic in particle state does not exist either due to the appearance of a big-bang singularity or horizon in the past. When both the in and out particle states exist, there may be a relation between the in-in and in-out formalisms. For instance, a simple relation between the asymptotic distribution functions of these two versions of QVE is given in \cite{Huet2014}. Notice that recent studies of NAQVE are in scalar QED. The NAQVE in spinor QED is still not derived. The instantaneous relations between the NAQVE and QVE, both in spinor and scalar QED, are also not established. What is the difference between the NAQVE in spinor and scalar QED is still unclear. Whether the out-of-phase behavior of momentum distributions between spinor and scalar QED for QVE, discovered in \cite{Hebenstreit2009} and explained in \cite{Dumlu2010}, still exists for NAQVE is not yet explored. In this paper, we will focus on investigating the above issues in detail and further deepen the understanding of adiabatic and nonadiabatic QVE and their applications in vacuum pair production.

The structure of this paper is as follows: In Sec. \ref{sec:two} the NAQVE in spinor QED is derived by a time-dependent Bogoliubov transformation. In Sec \ref{sec:three} the instantaneous relation between NAQVE and QVE is established by three methods: subsection \ref{A} is by an explicitly analytical expression; subsection \ref{B} is by the Dirac equation in the V gauge; subsection \ref{C} is by introducing a turn-off electric field. Section \ref{sec:four} is a summary and discussion. For convenience and completeness, the NAQVE and its instantaneous relation to QVE in scalar QED are derived in Appendix \ref{app}.

\section{Derivation of the nonadiabatic QVE in spinor QED}
\label{sec:two}

For a spatially homogeneous and time-dependent electric field in the temporal gauge $A_\mu(\mathbf{x},t)=(0,-\mathbf{A}(t))=(0,0,0,-A_z(t))$, Dirac equation becomes
\begin{equation}\label{eqn:DiracEquation10}
\big\{i\gamma^{0}\partial_t+i\bm{\gamma}\cdot[\bm{\nabla}-i q \mathbf{A}(t)]-m\big\}\Psi^A(\mathbf{x},t)=0,
\end{equation}
where $\partial_t$ represents the partial derivative with respect to time, $q$ and $m$ are the charge and mass of the particle, respectively.
Carrying on the Fourier transform of the Dirac field
\begin{equation}\label{eqn:FourierDeco0}
\Psi^A(\mathbf{x},t)=\int\frac{d^3k}{(2\pi)^3}\Psi^A_\mathbf{k}(t)e^{i\mathbf{k}\cdot\mathbf{x}},
\end{equation}
one can obtain the Fourier modes $\Psi^A_\mathbf{k}(t)$ satisfying
\begin{equation}\label{eqn:DiracEquation11}
\big[i\gamma^{0}\partial_t-\bm{\gamma}\cdot \mathbf{p}(t)-m \big]\Psi^A_\mathbf{k}(t)=0,
\end{equation}
where $\mathbf{k}$ is the canonical momentum and $\mathbf{p}(t)=\mathbf{k}-q\mathbf{A}(t)$ is the kinetic momentum. Assuming
\begin{equation}\label{eqn:DiracEquation12}
\Psi^A_\mathbf{k}(t)=\big[i\gamma^{0}\partial_t-\bm{\gamma}\cdot \mathbf{p}(t)+m\big]\psi^A_\mathbf{k}(t),
\end{equation}
and using Eq. (\ref{eqn:DiracEquation11}), we have
\begin{equation}\label{eqn:PDE1}
\Big[\partial^2_t+\omega_\mathbf{k}^2(t)+iqE_z(t)\gamma^0\gamma^3\Big]\psi^A_\mathbf{k}(t)=0,
\end{equation}
where $\omega_\mathbf{k}(t)=\{[\mathbf{k}-q\mathbf{A}(t)]^2+m^2\}^{1/2}$ is the total energy of particles and $E_z(t)=-dA_z(t)/dt$ is the electric field.
Using $R_s$, the eigenvectors of $\gamma^0\gamma^3$, to expand the functions $\psi^A_\mathbf{k}(t)$ as
\begin{equation}
\psi^A_\mathbf{k}(t)=\sum_{s=1}^4\chi_\mathbf{k}^s(t)R_s,
\end{equation}
where $R_1\!=\!\left(
                                   \begin{array}{cccc}
                                     1 & 0 & 0 & 0 \\
                                   \end{array}
                                 \right)^\textsf{T}$,
$R_2\!=\!\left(
                                   \begin{array}{cccc}
                                     0 & 0 & 0 & 1 \\
                                   \end{array}
                                 \right)^\textsf{T}$,
$R_3\!=\!\left(
                                   \begin{array}{cccc}
                                     0 & -1 & 0 & 0 \\
                                   \end{array}
                                 \right)^\textsf{T}$,
$R_4\!=\!\left(
                                   \begin{array}{cccc}
                                     0 & 0 & -1 & 0 \\
                                   \end{array}
                                 \right)^\textsf{T}$
in the chiral representation of $\gamma$-matrices, and $\gamma^0\gamma^3R_{s=\{1,2\}}=+1R_{s=\{1,2\}}$,
$\gamma^0\gamma^3R_{s=\{3,4\}}=-1R_{s=\{3,4\}}$, then Eq. (\ref{eqn:PDE1}) reads
\begin{equation}\label{ODEs}
\begin{split}
\big[\partial^2_t+\omega^2_\mathbf{k}(t)+iqE_z(t)\big]
\chi^{s=\{1,2\}}_\mathbf{k}(t)=0,\\
\big[\partial^2_t+\omega^2_\mathbf{k}(t)-iqE_z(t)\big]
\chi^{s=\{3,4\}}_\mathbf{k}(t)=0.
\end{split}
\end{equation}
These equations are overdetermined for the Dirac equation and the redundancy can be removed by only choosing $s=\{1,2\}$ or $s=\{3,4\}$. Here we choose the former case.

The two linearly independent solutions of Eq. (\ref{ODEs}), $\chi^+_\mathbf{k}(t)$ and $\chi^-_\mathbf{k}(t)$, can be used to expand the field operator $\hat{\Psi}^A(\mathbf{x},t)$ as
\begin{equation}\label{eqn:FieldOperator1}
\begin{split}
\hat{\Psi}^A(\mathbf{x},t)=\!\int\!\frac{d^3k}{(2\pi)^3}
\sum_{s=1}^2\big[&\hat{b}_{\mathbf{k},s}u_{\mathbf{k},s}(t)\\
&+\hat{d}_{-\mathbf{k},s}^{\dagger}v_{-\mathbf{k},s}(t)\big]
e^{i\mathbf{k}\cdot\mathbf{x}},
\end{split}
\end{equation}
where $\hat{b}_{\mathbf{k},s}$ and $\hat{d}_{-\mathbf{k},s}^{\dagger}$ are the time-independent annihilation and creation operators which both fulfil the standard fermionic anticommutation relations,
\begin{equation}\label{eqn:FieldOperator2}
\begin{split}
u_{\mathbf{k},s}(t)&=\big[i\gamma^{0}\partial_t-\bm{\gamma}\cdot \mathbf{p}(t)+m \big]\chi^+_\mathbf{k}(t)R_s,\\
v_{-\mathbf{k},s}(t)&=\big[i\gamma^{0}\partial_t-\bm{\gamma}\cdot \mathbf{p}(t)+m \big]\chi^-_\mathbf{k}(t)R_s,
\end{split}
\end{equation}
are the positive and negative energy states.
In the absence of the external electric field, the Dirac equation has the free plane wave solutions and the positive and negative energy states have the form of
\begin{equation}\label{eqn:FreeState1}
\begin{split}
\widetilde{u}_{\mathbf{k},s}(t)&=\big[\gamma^{0}\omega_\mathbf{k}(t_0)
-\bm{\gamma}\cdot \mathbf{p}(t_0)+m\big]\widetilde{\chi}^+_\mathbf{k}(t)R_s,\quad\\
\widetilde{v}_{-\mathbf{k},s}(t)&=\big[-\gamma^{0}\omega_\mathbf{k}(t_0)
-\bm{\gamma}\cdot \mathbf{p}(t_0)+m \big]\widetilde{\chi}^-_\mathbf{k}(t)R_s,
\end{split}
\end{equation}
where $t_0$ is the initial time, $\omega_\mathbf{k}(t_0)=[\mathbf{p}^2(t_0)+m^2]^{1/2}$, $\mathbf{p}(t_0)=\mathbf{k}-q\mathbf{A}(t_0)$, and
\begin{equation}\label{chit0}
\widetilde{\chi}^{\pm}_\mathbf{k}(t)=\frac{e^{\mp i\omega_\mathbf{k}(t_0)(t-t_0)}}{\sqrt{2\omega_\mathbf{k}(t_0)[\omega
_\mathbf{k}(t_0)\mp p_z(t_0)]}}.
\end{equation}
For the sake of intuitive, we show the explicit form of Eq. (\ref{eqn:FreeState1})
\begin{equation}\label{eqn:u1}
\widetilde{u}_{\mathbf{k},1}(t)=\left(
  \begin{array}{c}
   m \\
   0 \\
   -\omega_\mathbf{k}(t_0)+p_z(t_0) \\
   p_x(t_0)+ip_y(t_0) \\
  \end{array}
\right)\widetilde{\chi}^{+}_\mathbf{k}(t),
\end{equation}
\begin{equation}
\widetilde{u}_{\mathbf{k},2}(t)=\left(
  \begin{array}{c}
    -p_x(t_0)+ip_y(t_0) \\
    -\omega_\mathbf{k}(t_0)+p_z(t_0) \\
    0 \\
    m \\
  \end{array}
\right)\widetilde{\chi}^{+}_\mathbf{k}(t),
\end{equation}
\begin{equation}\label{eqn:v1}
\widetilde{v}_{-\mathbf{k},1}(t)=\left(
  \begin{array}{c}
   m \\
   0 \\
   \omega_\mathbf{k}(t_0)+p_z(t_0) \\
   p_x(t_0)+ip_y(t_0) \\
  \end{array}
\right)\widetilde{\chi}^{-}_\mathbf{k}(t),\;\,
\end{equation}
\begin{equation}
\widetilde{v}_{-\mathbf{k},2}(t)=\left(
  \begin{array}{c}
    -p_x(t_0)+ip_y(t_0) \\
    \omega_\mathbf{k}(t_0)+p_z(t_0) \\
    0 \\
    m \\
  \end{array}
\right)\widetilde{\chi}^{-}_\mathbf{k}(t).
\end{equation}
Since the plane wave solutions is also an orthogonal complete set for the Dirac field, the field operator can also be expanded as
\begin{equation}\label{eqn:FieldOperator3}
\begin{split}
\hat{\Psi}^A(\mathbf{x},t)=\!\!\int\!\frac{d^3k}{(2\pi)^3}
\sum_{s=1}^2\big[&\,\hat{b}_{\mathbf{k},s}(t)
\widetilde{u}_{\mathbf{k},s}(t)\\
&+\hat{d}_{-\mathbf{k},s}^{\,\dagger}(t)\widetilde{v}_{-\mathbf{k},s}(t)\big]
e^{i\mathbf{k}\cdot\mathbf{x}},
\end{split}
\end{equation}
where $\hat{b}_{\mathbf{k},s}(t)$ and $\hat{d}_{-\mathbf{k},s}^{\dagger}(t)$ are the time-dependent annihilation and creation operators satisfying the equal-time anticommutation relations.

The time-dependent annihilation and creation operators in Eq. (\ref{eqn:FieldOperator3}) can be associated with the time-independent ones in Eq. (\ref{eqn:FieldOperator1}) by a time-dependent Bogoliubov transformation
\begin{equation}\label{eqn:BogoliubovTransformation}
\begin{split}
\hat{b}_{\mathbf{k},s}(t)&=\alpha_\mathbf{k}(t)\hat{b}_{\mathbf{k},s}
-\beta^*_\mathbf{k}(t)\hat{d}^\dagger_{-\mathbf{k},s},\\
\hat{d}^{\,\dagger}_{-\mathbf{k},s}(t)&=\beta_\mathbf{k}(t)
\hat{b}_{\mathbf{k},s}+\alpha^*_\mathbf{k}(t)\hat{d}^\dagger_{-\mathbf{k},s},
\end{split}
\end{equation}
where
\begin{equation}\label{eqn:BogoliubovCoefficient}
\begin{split}
\alpha_\mathbf{k}(t)&=\widetilde{u}^{\,\dagger}_{\mathbf{k},s}(t)u_{\mathbf{k},s}(t)=
[\widetilde{v}^{\,\dagger}_{-\mathbf{k},s}(t)v_{-\mathbf{k},s}(t)]^*, \\
\beta_\mathbf{k}(t)&=\widetilde{v}^{\,\dagger}_{-\mathbf{k},s}(t)u_{\mathbf{k},s}(t)=
-[\widetilde{u}^{\,\dagger}_{\mathbf{k},s}(t)v_{-\mathbf{k},s}(t)]^*,
\end{split}
\end{equation}
are the Bogoliubov coefficients and satisfy $|\alpha_\mathbf{k}(t)|^2+|\beta_\mathbf{k}(t)|^2=1$ which
is obtained from the anticommutation relations of annihilation and creation operators. Substituting Eqs. (\ref{eqn:FieldOperator2}) and (\ref{eqn:FreeState1}) into Eq. (\ref{eqn:BogoliubovCoefficient}), the Bogoliubov coefficients can be expressed as
\begin{equation}\label{eqn:ab1}
\begin{split}
\alpha_\mathbf{k}(t)=\epsilon_\perp
\widetilde{\chi}^-_\mathbf{k}(t)\{&i\partial_t+\omega_\mathbf{k}(t_0)\\
&+[p_z(t_0)-p_z(t)]\}\chi^+_\mathbf{k}(t), \\
\beta_\mathbf{k}(t)=-\epsilon_\perp
\widetilde{\chi}^+_\mathbf{k}(t)\{&i\partial_t-\omega_\mathbf{k}(t_0)\\
&+[p_z(t_0)-p_z(t)]\}\chi^+_\mathbf{k}(t),
\end{split}
\end{equation}
where $\epsilon_\perp=[p_x^2(t_0)+p_y^2(t_0)+m^2]^{1/2}=(k_x^2+k_y^2+m^2)^{1/2}$.
In virtue of Eq. (\ref{ODEs}), the time derivatives of the Bogoliubov coefficients are
\begin{equation}\label{eqn:abdot1}
\begin{split}
\dot{\alpha}_\mathbf{k}(t)=&i\epsilon_\perp[p_z(t)-p_z(t_0)]
\widetilde{\chi}^-_\mathbf{k}(t)\{i\partial_t-\omega_\mathbf{k}(t_0) \\
&-[p_z(t)+p_z(t_0)]\}\chi^+_\mathbf{k}(t), \\
\dot{\beta}_\mathbf{k}(t)=&-i\epsilon_\perp[p_z(t)-p_z(t_0)]
\widetilde{\chi}^+_\mathbf{k}(t)\{i\partial_t+\omega_\mathbf{k}(t_0) \\
&-[p_z(t)+p_z(t_0)]\}\chi^+_\mathbf{k}(t).
\end{split}
\end{equation}
Using Eq. (\ref{eqn:ab1}), the above equations can be further written as
\begin{equation}\label{eqn:abdot2}
\begin{split}
\dot{\alpha}_\mathbf{k}(t)=&-iP_\mathbf{k}(t)\alpha_\mathbf{k}(t)
-iQ_\mathbf{k}(t)\beta_\mathbf{k}(t), \\
\dot{\beta}_\mathbf{k}(t)=&-iQ^*_\mathbf{k}(t)\alpha_\mathbf{k}(t)
+iP_\mathbf{k}(t)\beta_\mathbf{k}(t),
\end{split}
\end{equation}
where $P_\mathbf{k}(t)=p_z(t_0)[p_z(t)-p_z(t_0)]/\omega_\mathbf{k}(t_0)$ and $Q_\mathbf{k}(t)=\epsilon_\perp[p_z(t)-p_z(t_0)]
e^{2i\omega_\mathbf{k}(t_0)(t-t_0)}/\omega_\mathbf{k}(t_0)$.
Then let $\widetilde{\alpha}_\mathbf{k}(t)=\alpha_\mathbf{k}(t)
e^{i\int^t_{t_0}P_\mathbf{k}(\tau)d\tau}$ and $\widetilde{\beta}_\mathbf{k}(t)=\beta_\mathbf{k}(t)
e^{-i\int^t_{t_0}P_\mathbf{k}(\tau)d\tau}$, we have
\begin{equation}\label{eqn:abdot3}
\begin{split}
\dot{\widetilde{\alpha}}_\mathbf{k}(t)=&\frac{i}{2}Q^+_\mathbf{k}(t)
\widetilde{\beta}_\mathbf{k}(t)e^{2i\int^t_{t_0}\Omega^+_\mathbf{k}(\tau)d\tau}, \\
\dot{\widetilde{\beta}}_\mathbf{k}(t)=&\frac{i}{2}Q^+_\mathbf{k}(t)
\widetilde{\alpha}_\mathbf{k}(t)e^{-2i\int^t_{t_0}\Omega^+_\mathbf{k}(\tau)d\tau},
\end{split}
\end{equation}
where $Q^+_\mathbf{k}(t)=2\epsilon_\perp[p_z(t_0)-p_z(t)]/\omega_\mathbf{k}(t_0)$ and $\Omega^+_\mathbf{k}(\tau)=[\epsilon_\perp^2+p_z(t_0)p_z(t)]
/\omega_\mathbf{k}(t_0)$.

The momentum distribution function of particle pairs produced from the initial vacuum by the electric field for a specific spin is defined as
\begin{equation}\label{eqn:MS}
\begin{split}
f^+_{\mathbf{k},s}(t)\equiv&\lim_{V\rightarrow\infty}\frac{\langle \mathrm{vac}|\hat{b}^{\,\dagger}_{\mathbf{k},s}(t)
\hat{b}_{\mathbf{k},s}(t)|\mathrm{vac}\rangle}{V}\\
=&|\beta_\mathbf{k}(t)|^2=|\widetilde{\beta}_\mathbf{k}(t)|^2,
\end{split}
\end{equation}
where $V$ is the configuration space volume, whose appearance is for  canceling the divergence in the anticommutators of time-independent annihilation and creation operators, $|\mathrm{vac}\rangle$ is the vacuum state satisfying $\hat{b}_{\mathbf{k},s}|\mathrm{vac}\rangle=0$ and $\hat{d}_{-\mathbf{k},s}|\mathrm{vac}\rangle=0$. By introducing the quantity $c_\mathbf{k}(t)=\widetilde{\alpha}^*_\mathbf{k}(t)
\widetilde{\beta}_\mathbf{k}(t)$,
we can obtain the nonadiabatic quantum Vlasov equation (NAQVE) in integral-differential form:
\begin{equation}\label{eqn:QKT}
\begin{split}
\dot{f}^+_\mathbf{k}(t)=\frac{1}{2}Q^+_\mathbf{k}(t)
\int_{t_0}^t\!dt'&Q^+_\mathbf{k}(t')[1-2f^+_\mathbf{k}(t')]\\
&\times\cos[2\theta^+_\mathbf{k}(t',t)],
\end{split}
\end{equation}
where $\theta^+_\mathbf{k}(t',t)=\int^t_{t'}\Omega^+_\mathbf{k}(\tau)d\tau$. Since the particles with different spin states in the absence of magnetic field behave identically, the subscript $s$ is omitted for convenience. Note that for the adiabatic QVE the quantities $Q^+_\mathbf{k}(t)=eE(t)\epsilon_\perp/\omega^2_\mathbf{k}(t)$ and $\theta^+_\mathbf{k}(t',t)=\int^t_{t'}\omega_\mathbf{k}(\tau)d\tau$. Introducing two auxiliary quantities
\begin{equation}
g^+_\mathbf{k}(t)=\!\int_{t_0}^t d t'\,Q^+_\mathbf{k}(t')[1-2f^+_\mathbf{k}(t')]
\cos[2\theta^+_\mathbf{k}(t',t)]\nonumber
\end{equation}
and
\begin{equation}
h^+_\mathbf{k}(t)=\!\int_{t_0}^t d t'\,Q^+_\mathbf{k}(t')[1-2f^+_\mathbf{k}(t')]
\sin[2\theta^+_\mathbf{k}(t',t)],\nonumber
\end{equation}
equation (\ref{eqn:QKT}) can be equivalently transformed into a set of ODEs:
\begin{eqnarray}\label{QKE2}
\dot{f}^+_\mathbf{k}(t)&\!=\!&\frac{1}{2}Q^+_\mathbf{k}(t)g^+_\mathbf{k}(t), \nonumber\\
\dot{g}^+_\mathbf{k}(t)&\!=\!&Q^+_\mathbf{k}(t)[1-2f^+_\mathbf{k}(t)]
-2\,\Omega^+_\mathbf{k}(t)h^+_\mathbf{k}(t)  , \qquad\;\\
\dot{h}^+_\mathbf{k}(t)&\!=\!&2\,\Omega^+_\mathbf{k}(t)g^+_\mathbf{k}(t),\nonumber
\end{eqnarray}
with the initial conditions $f^+_\mathbf{k}(t_0)=g^+_\mathbf{k}(t_0)=h^+_\mathbf{k}(t_0)=0$.

\section{Relation between the NAQVE and QVE in spinor QED}
\label{sec:three}

In this section, we will show the difference between the NAQVE and the traditional (or adiabatic) QVE in spinor QED, and provide three ways to establish the relation between them.

\begin{figure}[!ht]
\centering
\includegraphics[width=0.45\textwidth]{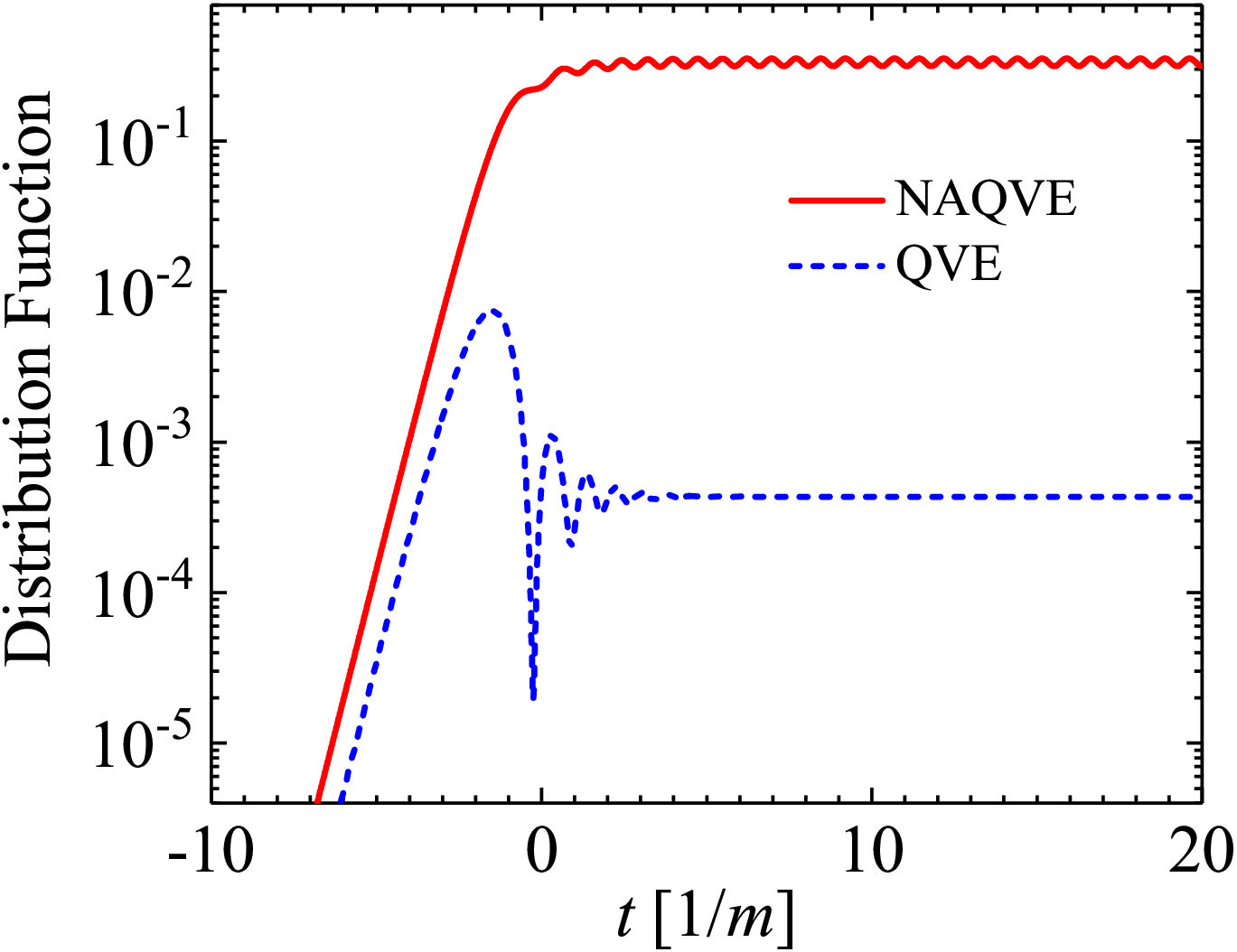}%
\caption{Time evolution of the distribution functions for NAQVE and QVE in the case of $k_x=0$ and $k_y=k_z=0.1m$. The parameters of the field (\ref{eqn:electricfield}) are $E_0=1E_{\rm{cr}}$ and $\tau=2/m$.
\label{fig:Fig1}}
\end{figure}

To see the difference between the NAQVE and the QVE, we depict the time evolution of the distribution functions $f^+_\mathbf{k}(t)$ with $k_x=0$ and $k_y=k_z=0.1m$ for NAQVE and QVE in Fig. \ref{fig:Fig1}. The electric field is
\begin{equation}\label{eqn:electricfield}
E_z(t)=E_0\mathrm{sech}^2(t/\tau),
\end{equation}
where $E_0=1E_{\mathrm{cr}}$ is the field amplitude and $\tau=2/m$ denotes the pulse duration. The corresponding vector potential is $A_z(t)=-E_0\tau[1+\mathrm{tanh}(t/\tau)]$. From the figure, one can see that the distribution functions for NAQVE and QVE are very different. Especially when the electric field is turned off, there is a difference of nearly three orders of magnitude between them. The distribution function for the NAQVE is nonconvergent and oscillates over time. The main reason for this difference is that the definition of the vacuum state for the NAQVE and QVE is different. For the NAQVE, the vacuum state corresponding to the annihilation and creation operators of nonadiabatic particles is a vacuum state in which the vacuum energy is time-independent. For the QVE, however, the vacuum state corresponding to the annihilation and creation operators of adiabatic particles is a vacuum state in which the vacuum energy is time-dependent. Even so, these two versions of QVE can be related to each other.

\subsection{Relate NAQVE to QVE by an explicitly analytical expression}\label{A}

Since there is no difference between the different spin states of particles in a linearly polarized electric field, we only consider one of the spin states for convenience. The Fourier mode operator can be expanded as
\begin{equation}\label{eqn:Expansion1}
\begin{split}
\hat{\Psi}^A_{\mathbf{k},1}(t)=&\hat{b}_{\mathbf{k},1}(t)
\widetilde{u}_{\mathbf{k},1}(t)+\hat{d}_{-\mathbf{k},1}^{\dagger}(t)
\widetilde{v}_{-\mathbf{k},1}(t)\\
=&\hat{B}_{\mathbf{k},1}(t)U_{\mathbf{k},1}(t)
+\hat{D}_{-\mathbf{k},1}^{\dagger}(t)V_{-\mathbf{k},1}(t),
\end{split}
\end{equation}
where $\hat{B}_{\mathbf{k},1}(t)$, $\hat{D}_{-\mathbf{k},1}^{\dagger}(t)$ are the annihilation and creation operators of adiabatic particles, $U_{\mathbf{k},1}(t)$, $V_{-\mathbf{k},1}(t)$ are the adiabatic basis and their explicit form are
\begin{equation}\label{eqn:U1}
U_{\mathbf{k},1}(t)=\left(
  \begin{array}{c}
   m \\
   0 \\
   -\omega_\mathbf{k}(t)+p_z(t) \\
   p_x(t)+ip_y(t) \\
  \end{array}
\right)\widetilde{X}^{+}_\mathbf{k}(t),
\end{equation}
\begin{equation}\label{eqn:V1}
V_{-\mathbf{k},1}(t)=\left(
  \begin{array}{c}
   m \\
   0 \\
   \omega_\mathbf{k}(t)+p_z(t) \\
   p_x(t)+ip_y(t) \\
  \end{array}
\right)\widetilde{X}^{-}_\mathbf{k}(t),\;\;
\end{equation}
where
\begin{equation}\label{eqn:phase}
\widetilde{X}^{\pm}_\mathbf{k}(t)=\frac{e^{\mp i\int^t_{t_0}
\omega_\mathbf{k}(\tau)d\tau}}{\sqrt{2\omega_\mathbf{k}(t)
[\omega_\mathbf{k}(t)\mp p_z(t)]}},
\end{equation}
$p_x(t)=p_x(t_0)=k_x$ and $p_y(t)=p_y(t_0)=k_y$ for the linearly polarized electric field (\ref{eqn:electricfield}).
Then, the annihilation and creation operators of nonadiabatic particles can be expressed by the ones of adiabatic particles
\begin{equation}\label{eqn:Expansion2}
\begin{split}
\hat{b}_{\mathbf{k},1}(t)=&\hat{B}_{\mathbf{k},1}(t)
[\widetilde{u}^\dagger_{\mathbf{k},1}(t)U_{\mathbf{k},1}(t)]\\
&+\hat{D}_{-\mathbf{k},1}^{\dagger}(t)
[\widetilde{u}^\dagger_{\mathbf{k},1}(t)V_{-\mathbf{k},1}(t)],\\
\hat{b}^\dagger_{\mathbf{k},1}(t)=&\hat{B}^\dagger_{\mathbf{k},1}(t)
[\widetilde{u}^\dagger_{\mathbf{k},1}(t)U_{\mathbf{k},1}(t)]^\dagger\\
&+\hat{D}_{-\mathbf{k},1}(t)
[\widetilde{u}^\dagger_{\mathbf{k},1}(t)V_{-\mathbf{k},1}(t)]^\dagger.
\end{split}
\end{equation}
According to the above equation and the definition Eq. (\ref{eqn:MS}), we find
\begin{equation}\label{eqn:fvsF}
\begin{split}
f^+_{\mathbf{k},1}(t)=&|\widetilde{u}^\dagger_{\mathbf{k},1}(t)
U_{\mathbf{k},1}(t)|^2F^+_{\mathbf{k},1}(t)\\
&\!-\!2\Re\{[\widetilde{u}^\dagger_{\mathbf{k},1}(t)
U_{\mathbf{k},1}(t)]^\dagger[\widetilde{u}^\dagger_{\mathbf{k},1}(t)
V_{\!-\mathbf{k},1}(t)]C^+_{\mathbf{k},1}(t)\}\\
&+|\widetilde{u}^\dagger_{\mathbf{k},1}(t)V_{-\mathbf{k},1}(t)|^2
[1-F^+_{\mathbf{k},1}(t)],
\end{split}
\end{equation}
where $F^+_{\mathbf{k},1}(t)=\lim_{V\rightarrow\infty}\frac{\langle \mathrm{vac}|\hat{B}^{\,\dagger}_{\mathbf{k},1}(t)
\hat{B}_{\mathbf{k},1}(t)|\mathrm{vac}\rangle}{V}$ is the momentum distribution function of created adiabatic fermions, $C^+_{\mathbf{k},1}(t)=\lim_{V\rightarrow\infty}\frac{\langle \mathrm{vac}|\hat{D}^{\,\dagger}_{-\mathbf{k},1}(t)
\hat{B}^{\,\dagger}_{\mathbf{k},1}(t)
|\mathrm{vac}\rangle}{V}$ is the time-dependent pair correlation function describing the production of a quasi-fermion pair, and $\Re\{\}$ denotes the real part of a complex number.

From Eqs. (\ref{eqn:u1}), (\ref{eqn:v1}), (\ref{eqn:U1}), (\ref{eqn:V1}), one can obtain
\begin{equation}\label{eqn:uU}
\begin{split}
\widetilde{u}^\dagger_{\mathbf{k},1}(t)U_{\mathbf{k},1}(t)=&
\sqrt{\frac{\omega_\mathbf{k}(t_0)-p_z(t_0)}{4\omega_\mathbf{k}(t_0)
\omega_\mathbf{k}(t)[\omega_\mathbf{k}(t)-p_z(t)]}}\\
&\times[\omega_\mathbf{k}(t_0)\!+p_z(t_0)\!+\omega_\mathbf{k}(t)\!-p_z(t)]\\
&\times e^{-i\int_{t_0}^{t}[\omega_\mathbf{k}(\tau)-\omega_\mathbf{k}(t_0)]d\tau},
\end{split}
\end{equation}
\begin{equation}\label{eqn:uV}
\begin{split}
\widetilde{u}^\dagger_{\mathbf{k},1}(t)V_{\!-\mathbf{k},1}(t)=&
\sqrt{\frac{\omega_\mathbf{k}(t_0)-p_z(t_0)}{4\omega_\mathbf{k}(t_0)
\omega_\mathbf{k}(t)[\omega_\mathbf{k}(t)+p_z(t)]}}\\
&\times[\omega_\mathbf{k}(t_0)\!+\!p_z(t_0)\!-\!\omega_\mathbf{k}(t)\!-\!p_z(t)]\\
&\times e^{i\int_{t_0}^{t}[\omega_\mathbf{k}(\tau)+\omega_\mathbf{k}(t_0)]d\tau}.
\end{split}
\end{equation}
Then
\begin{equation}\label{eqn:uU2}
\begin{split}
|\widetilde{u}^\dagger_{\mathbf{k},1}(t)U_{\mathbf{k},1}(t)|^2=&
\frac{\epsilon^2_\perp+p_z(t_0)p_z(t)}{2\omega_\mathbf{k}(t_0)
\omega_\mathbf{k}(t)}+\frac{1}{2},\\
|\widetilde{u}^\dagger_{\mathbf{k},1}(t)V_{-\mathbf{k},1}(t)|^2=&
-\frac{\epsilon^2_\perp+p_z(t_0)p_z(t)}{2\omega_\mathbf{k}(t_0)
\omega_\mathbf{k}(t)}+\frac{1}{2},
\end{split}
\end{equation}
and
\begin{equation}\label{eqn:uUuV}
\begin{split}
&[\widetilde{u}^\dagger_{\mathbf{k},1}(t)U_{\mathbf{k},1}(t)]^\dagger
[\widetilde{u}^\dagger_{\mathbf{k},1}(t)V_{-\mathbf{k},1}(t)]\\
=&
\frac{\epsilon_\perp[p_z(t_0)-p_z(t)]}{2\omega_\mathbf{k}(t_0)
\omega_\mathbf{k}(t)}e^{2i\int_{t_0}^{t}\omega_\mathbf{k}(\tau)d\tau}.
\end{split}
\end{equation}
Finally, equation (\ref{eqn:fvsF}) becomes
\begin{equation}\label{eqn:Relation1}
\begin{split}
f^+_{\mathbf{k},1}(t)=&\frac{\epsilon^2_\perp+p_z(t_0)p_z(t)}{2\omega_\mathbf{k}(t_0)
\omega_\mathbf{k}(t)}[2F^+_{\mathbf{k},1}(t)-1]+\frac{1}{2}\\
&-\frac{\epsilon_\perp[p_z(t_0)-p_z(t)]}{\omega_\mathbf{k}(t_0)
\omega_\mathbf{k}(t)}\Re\Big\{C^+_{\mathbf{k},1}(t)
e^{2i\int_{t_0}^{t}\omega_\mathbf{k}(\tau)d\tau}\Big\},
\end{split}
\end{equation}
where
\begin{equation}\label{eqn:Ck}
\begin{split}
C^+_{\mathbf{k},1}(t)=\!\int_{t_0}^{t}\!dt'&\frac{qE_z(t')\epsilon_\perp}
{2\omega_\mathbf{k}^2(t')}[2F^+_{\mathbf{k},1}(t')-1]\\
&\times e^{-2i\int_{t_0}^{t'}\omega_\mathbf{k}(\tau)d\tau}.
\end{split}
\end{equation}
According to the integro-differential form of adiabatic QVE,
\begin{equation}\label{eqn:CkExp}
\begin{split}
&\Re\Big\{C^+_{\mathbf{k},1}(t)e^{2i\int_{t_0}^{t}\omega_\mathbf{k}(\tau)d\tau}\Big\}\\
=&\!\int_{t_0}^{t}dt'\frac{qE_z(t')\epsilon_\perp}{2\omega_\mathbf{k}^2(t')}
[2F^+_{\mathbf{k},1}(t')\!-\!1]\cos\Big[2\!\int_{t'}^{t}\!\omega_\mathbf{k}(\tau)d\tau\Big]\\
=&-\frac{\omega_\mathbf{k}^2(t)}{qE_z(t)\epsilon_\perp}\dot{F}^+_{\mathbf{k},1}(t).
\end{split}
\end{equation}
Thus, equation (\ref{eqn:Relation1}) can be rewritten as 
\begin{equation}\label{eqn:Relation1'}
\begin{split}
f^+_{\mathbf{k},1}(t)=&\frac{\epsilon^2_\perp+p_z(t_0)p_z(t)}{2\omega_\mathbf{k}(t_0)
\omega_\mathbf{k}(t)}[2F^+_{\mathbf{k},1}(t)-1]+\frac{1}{2}\\
&+\frac{p_z(t_0)-p_z(t)}{\dot{p}_z(t)}
\frac{\omega_\mathbf{k}(t)}{\omega_\mathbf{k}(t_0)}\dot{F}^+_{\mathbf{k},1}(t).
\end{split}
\end{equation}
This equation gives us the specific relation between the NAQVE and the QVE.
It can be seen that, generally, these two versions of QVE are different. In order to let them give the same result at the final time $t_f$ when the electric field is turned off, it should ensure $p_z(t_f)=p_z(t_0)$, i.e., $A_z(t_f)=A_z(t_0)$. It is worth noting that although the electric field is turned off, the vector potential may not be zero and still affect pair production. This reflects the nonlocal nature of quantum mechanics. For instance, in Fig. \ref{fig:Fig1}, when the electric field is turned off the vector potential is $-4m/q$ and different from the initial value $0$. Then one can see that the distribution functions for NAQVE and QVE are different. 

\subsection{Relate NAQVE to QVE by the Dirac equation in the V gauge}\label{B}

From Eq. (\ref{eqn:BogoliubovCoefficient}) and Eq. (\ref{eqn:MS}), it can be seen that the NAQVE can be reproduced by projecting the negative energy solutions of the Dirac equation in the temporal gauge onto the positive-energy plane-wave solutions with a specific momentum in Eq. (\ref{eqn:FreeState1}) and calculating the square of its modulus. In this subsection, we will show that the NAQVE can turn into the QVE by replacing the negative energy solutions of the Dirac equation in the temporal gauge with those in the V gauge. The V gauge denotes that the four-vector potential only has the time component but gives the same electric field as in the temporal gauge.

The four-vector potential of the uniform and time-varying electric field in the V gauge has the form $A_\mu(\mathbf{x},t)=(\int^z_{z_0} \frac{\partial A_z(t)}{\partial t}dz', \mathbf{0})$, where $z_0$ is the zero point of electric potential. Then the Dirac equation (\ref{eqn:DiracEquation10}) becomes
\begin{equation}\label{eqn:ScalarGauge}
\Big\{i\gamma^{0}\Big[\partial_t+i q \int^z_{z_0} \frac{\partial A_z(t)}{\partial t}dz'\Big]+i\bm{\gamma}\cdot\bm{\nabla}
-m\Big\}\Psi^V(\mathbf{x},t)=0.
\end{equation}

Set
\begin{equation}
\Psi^V(\mathbf{x},t)=\int\frac{d^3k}{(2\pi)^3}\Psi^V_\mathbf{k}(t)
e^{i\mathbf{k}\cdot\mathbf{x}-iq\int_{z_0}^{z}A_z(t)dz'}
\end{equation}
and substitute it into Eq. (\ref{eqn:ScalarGauge}), one can obtain
\begin{equation}\label{eqn:ScalarGauge1}
\big[i\gamma^{0}\partial_t-\bm{\gamma}\cdot \mathbf{p}(t)-m \big]\Psi^V_\mathbf{k}(t)=0,
\end{equation}
Compared with Eq. (\ref{eqn:DiracEquation11}), it is easy to find that $\Psi^V_\mathbf{k}(t)=\Psi^A_\mathbf{k}(t)$ and $\Psi^V(\mathbf{x},t)=\Psi^A(\mathbf{x},t)e^{-iq\int_{z_0}^{z}A_z(t)dz'}$.
Therefore, the field operator $\hat{\Psi}^V(\mathbf{x},t)$ can be expanded as
\begin{equation}\label{eqn:Expand0}
\begin{split}
\hat{\Psi}^V(\mathbf{x},t)=\!\int\!&\frac{d^3k}{(2\pi)^3}
\sum_{s=1}^2\big[\hat{b}_{\mathbf{k},s}u_{\mathbf{k},s}(t)\\
&+\hat{d}_{-\mathbf{k},s}^{\dagger}v_{-\mathbf{k},s}(t)\big]
e^{i\mathbf{k}\cdot\mathbf{x}-iq\int_{z_0}^{z}A_z(t)dz'}.
\end{split}
\end{equation}

At zero electric field, equation (\ref{eqn:ScalarGauge}) has plane wave solutions $\Psi^V_\mathrm{free}(\mathbf{x},t)$, which are different from the plane wave solutions of Eq. (\ref{eqn:DiracEquation10}) only by a phase factor, i.e., $\Psi^V_\mathrm{free}(\mathbf{x},t)=
\Psi^A_\mathrm{free}(\mathbf{x},t)e^{-iq\int_{z_0}^{z}A_z(t_0)dz'}$. In particular, when the vector potential is zero at the initial time, these two plane wave solutions are the same. Since the plane wave solutions $\Psi^V_\mathrm{free}(\mathbf{x},t)$ are orthogonal and complete, the field operator can also be expressed as
\begin{equation}\label{eqn:Expand1}
\begin{split}
\hat{\Psi}^V(\mathbf{x},t)=&\!\int\!\frac{d^3k}{(2\pi)^3}
\sum_{s=1}^2\big[\hat{\mathfrak{b}}_{\mathbf{k},s}(t)\widetilde{u}_{\mathbf{k},s}(t)\\
&+\hat{\mathfrak{d}}_{-\mathbf{k},s}^{\dagger}(t)\widetilde{v}_{-\mathbf{k},s}(t)\big]
e^{i\mathbf{k}\cdot\mathbf{x}-iq\int_{z_0}^{z}A_z(t_0)dz'},
\end{split}
\end{equation}
where $\hat{\mathfrak{b}}_{\mathbf{k},s}(t)$ and $\hat{\mathfrak{d}}_{-\mathbf{k},s}^{\dagger}(t)$ are the time-dependent annihilation and creation operators, which are different from those in Eq. (\ref{eqn:FieldOperator3}).

According to Eqs. (\ref{eqn:Expand0}) and (\ref{eqn:Expand1}), we can obtain
\begin{equation}\label{eqn:AOperator}
\begin{split}
&\hat{\mathfrak{b}}_{\mathbf{k'},s}(t)=e^{iq[A_z(t)-A_z(t_0)]z_0}
\big[\hat{b}_{\mathbf{k},s}(t)
\widetilde{u}^\dagger_{\mathbf{k'},s}(t)u_{\mathbf{k},s}(t)\\
&\hspace{0.2cm}+\hat{d}^\dagger_{-\mathbf{k},s}(t)\widetilde{u}^\dagger_{\mathbf{k'},s}(t)
v_{-\mathbf{k},s}(t)\big]\big|_{\mathbf{k}\rightarrow\mathbf{k'}
+q[A_z(t)-A_z(t_0)]\mathbf{e}_z}
\end{split}
\end{equation}
and the momentum distribution function
\begin{equation}\label{eqn:MSV}
\begin{split}
\mathfrak{f}_{\,\mathbf{k'},s}(t)\equiv&\lim_{V\rightarrow\infty}\frac{\langle \mathrm{vac}|\hat{\mathfrak{b}}^{\,\dagger}_{\mathbf{k'},s}(t)
\hat{\mathfrak{b}}_{\mathbf{k'},s}(t)|\mathrm{vac}\rangle}{V}\\
=&|\widetilde{u}^\dagger_{\mathbf{k'},s}(t)
v_{-\{\mathbf{k'}+q[A_z(t)-A_z(t_0)]\mathbf{e}_z\},s}(t)\big|^2
\end{split}
\end{equation}
or
\begin{equation}\label{eqn:MSV1}
\begin{split}
&\mathfrak{f}_{\,\mathbf{k}-q[A_z(t)-A_z(t_0)]\mathbf{e}_z,s}(t)\\
=&|\widetilde{u}^\dagger_{\mathbf{k}-q[A_z(t)-A_z(t_0)]\mathbf{e}_z,s}(t)
v_{-\mathbf{k},s}(t)\big|^2,
\end{split}
\end{equation}
where $\mathbf{e}_z$ is the unit vector in the $z$ direction.

\begin{figure}[!ht]
\centering
\includegraphics[width=0.45\textwidth]{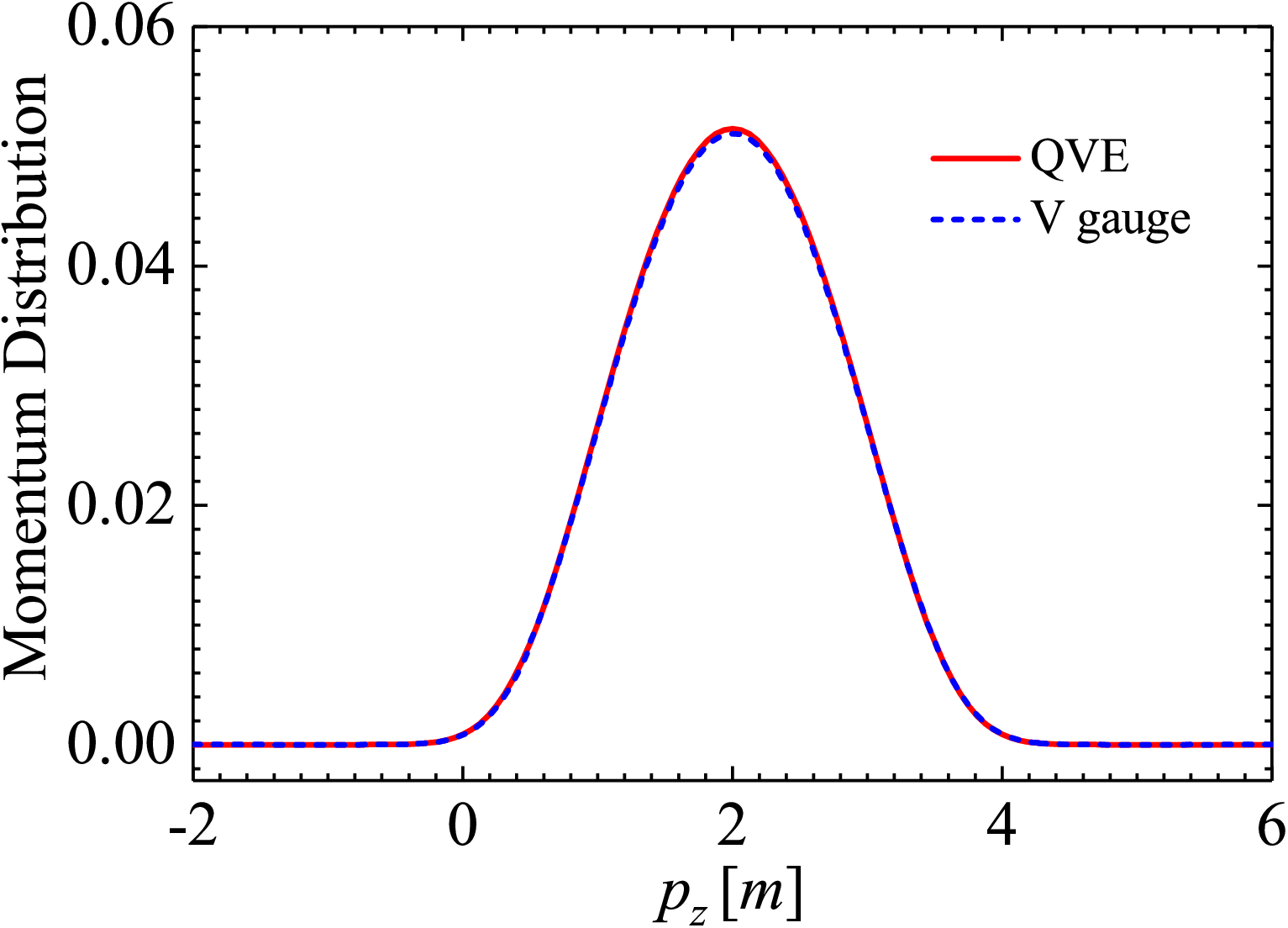}%
\caption{Comparison of the momentum distributions between the V gauge and the QVE at infinite time for $k_x=k_y=0$. The parameters of the field (\ref{eqn:electricfield}) are $E_0=1E_{\rm{cr}}$ and $\tau=2/m$.
\label{fig:Fig2}}
\end{figure}

From the right-hand side of the equation (\ref{eqn:MSV1}), one can see that $\widetilde{u}^\dagger_{\mathbf{k}-q[A_z(t)-A_z(t_0)]\mathbf{e}_z,s}(t)$ is the adiabatic basis, and the modulus squared term is nothing but the momentum distribution function of created adiabatic particles. Accordingly, equation (\ref{eqn:MSV}) gives us the kinetic momentum distribution. This indicates that one can reproduce the result of QVE by projecting the negative energy solutions of the Dirac equation in the V gauge, $\int\frac{d^3k}{(2\pi)^3}v_{-\mathbf{k},s}(t)
e^{i\mathbf{k}\cdot\mathbf{x}-iq\int_{z_0}^{z}[A_z(t)-A_z(t_0)]dz'}$, onto the positive-energy plane-wave solutions with a specific momentum in Eq. (\ref{eqn:FreeState1}) and calculating the square of its modulus. Furthermore, the negative energy solutions of the Dirac equation in the V gauge can be obtained numerically by evolving the negative-energy plane-wave solutions in Eq. (\ref{eqn:FreeState1}) using the split-operator technique \cite{Braun1999,Mocken20041,Mocken20042}. Figure \ref{fig:Fig2} shows the comparison of the kinetic momentum distributions between the V gauge and the QVE. It can be seen that they agree with each other very well. The difference between them primarily arises from the substitution of a locally uniform electric field for a spatially homogeneous one in the V gauge, which can be reduced by increasing the spatial scale of the local field. 

\subsection{Relate NAQVE to QVE by introducing a turn-off electric field}\label{C}

From Eq. (\ref{eqn:fvsF}), it can be seen that the NQVE can become the QVE when the vector potential at the final time equals the initial one. However, the electric field (\ref{eqn:electricfield}) does not satisfy this condition. To still achieve the change from NAQVE to QVE, we can introduce a turn-off electric field that does not affect the pair production in the original electric field. This requires that the turn-off field be weak enough and vary slowly with time. Moreover, the turn-off field should be introduced after the original electric field is turned off.

The form of the turn-off electric field is
\begin{equation}\label{eqn:TFEF}
E_{fz}(t)=-E_{f0}\mathrm{sech}^2((t-T_f)/\tau_f),
\end{equation}
where $E_{f0}$ is the field amplitude, $\tau_f$ denotes the pulse duration, and $T_f$ is the time delay relative to the electric field (\ref{eqn:electricfield}). So the total electric field is
\begin{equation}\label{eqn:TotalField}
E_{\rm{tot}}(t)=E_{0}\mathrm{sech}^2(t/\tau)-E_{f0}\mathrm{sech}^2((t-T_f)/\tau_f).
\end{equation}
To avoid the effect on the pair production in the original electric field, the parameters of the turn-off field should satisfy $E_{f0}\tau_f=E_0\tau$ and $T_f\geq10(\tau+\tau_f)$.

\begin{figure}[!ht]
\centering
\includegraphics[width=0.45\textwidth]{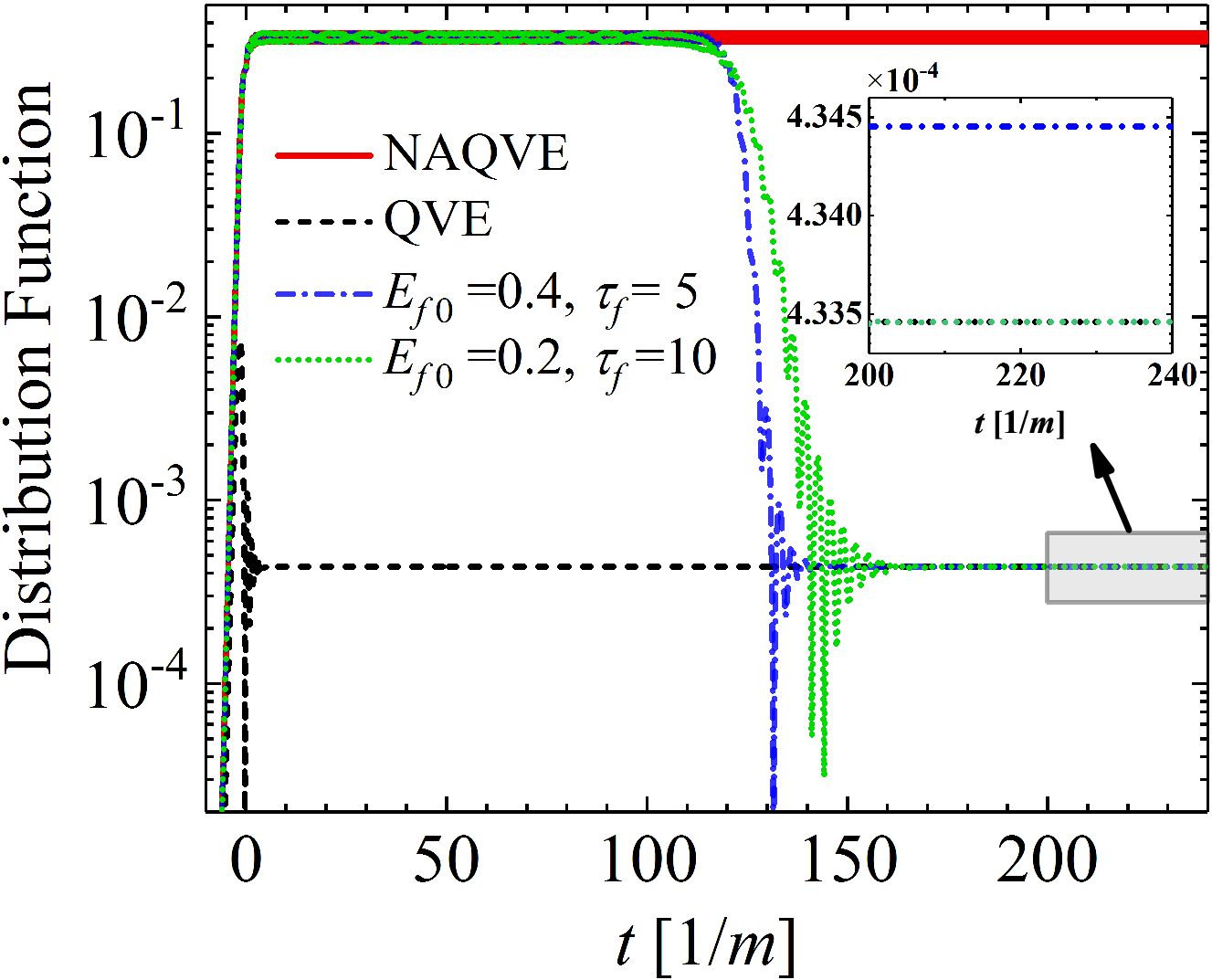}%
\caption{Time evolution of the distribution functions for NAQVE, QVE, and the NAQVE with a turn-off electric field in the case of $k_x=0$ and $k_y=k_z=0.1m$. The parameters of the turn-off field are $E_{f0}=0.4E_{\rm{cr}}, \tau_f=5/m, T_f=120/m$ (dash-dotted blue line) and $E_{f0}=0.2E_{\rm{cr}}, \tau_f=10/m, T_f=120/m$ (dotted green line), respectively. The parameters of the field (\ref{eqn:electricfield}) are $E_0=1E_{\rm{cr}}$ and $\tau=2/m$.
\label{fig:Fig3}}
\end{figure}

The time evolution of the distribution functions for NAQVE, QVE, and the NAQVE with a turn-off electric field are shown in Fig. \ref{fig:Fig3}. One can see that when $E_{f0}\leq0.2E_{\rm{cr}}$ (correspondingly $\tau_f\geq10/m$) and the original electric field is turned off, the NAQVE can give the same result as the QVE.

\section{Comparison of the NAQVE in spinor and scalar QED}
\label{sec:four}

\begin{figure}[!ht]
\centering
\includegraphics[width=0.45\textwidth]{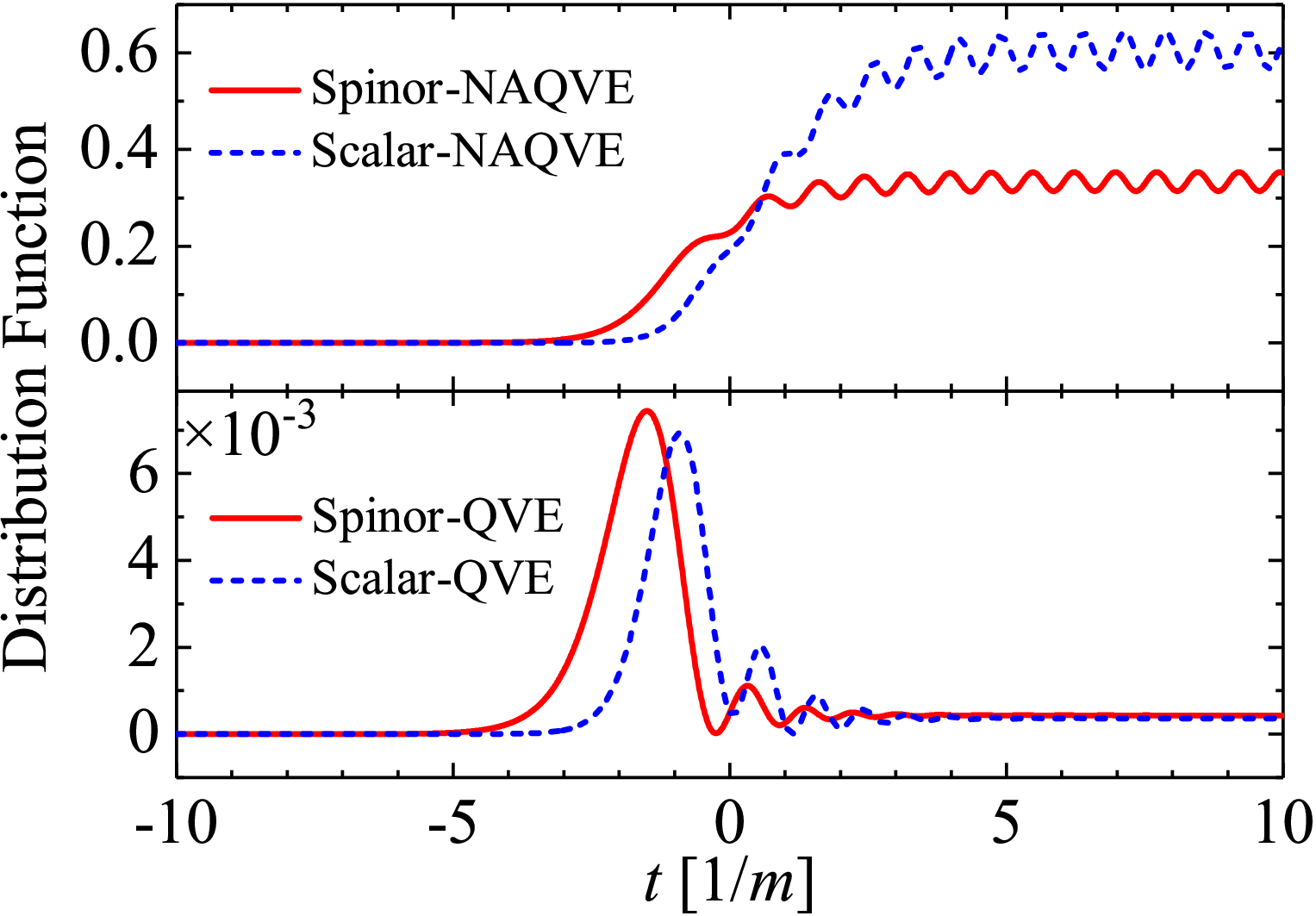}%
\caption{Time evolution of the distribution functions for NAQVE (upper panel) and QVE (lower panel) in the spinor and scalar QED in the case of $k_x=0$ and $k_y=k_z=0.1m$. The parameters of the field (\ref{eqn:electricfield}) are $E_0=1E_{\rm{cr}}$ and $\tau=2/m$.
\label{fig:Fig4}}
\end{figure}

From Eqs. (\ref{eqn:QKT}) and (\ref{QVE1}), one can see that there are three obvious differences between the NAQVE in spinor and scalar QED. The first one is the quantum statistical factor: similar to the QVE, there is a Pauli blocking factor $[1-2f^+_\mathbf{k}(t')]$ in spinor QED and a Bose enhancement factor $[1+2f^-_\mathbf{k}(t')]$ in scalar QED. However, these factors are more significant than those in QVE for the same field parameters because, in NAQVE, the distribution function may be close to or even greater than $1$ and cannot be neglected; see the upper panel in Fig. \ref{fig:Fig4}. The second one is the transition amplitude: $Q^+_\mathbf{k}(t)=\frac{2[p_z(t_0)-p_z(t)]}{\omega_\mathbf{k}(t_0)}\epsilon_\perp$ in spinor QED and $Q^-_\mathbf{k}(t)=\frac{\omega^2_\mathbf{k}(t_0)-\omega^2_\mathbf{k}(t)
}{\omega_\mathbf{k}(t_0)}=\frac{2[p_z(t_0)-p_z(t)]}{\omega_\mathbf{k}(t_0)
}\{[p_z(t_0)+p_z(t)]/2\}$ in scalar QED. The third difference is the dynamical phase: $\theta^+_\mathbf{k}(t',t)=\int_{t'}^t\frac{\epsilon_\perp^2+p_z(t_0)p_z(\tau)}{
\omega_\mathbf{k}(t_0)}d\tau$ in spinor QED and $\theta^-_\mathbf{k}(t',t)=\int_{t'}^t\frac{\omega^2_\mathbf{k}(t_0)
+\omega^2_\mathbf{k}(\tau)}{2\omega_\mathbf{k}(t_0)}d\tau=
\int_{t'}^t\frac{\epsilon_\perp^2+[p^2_z(t_0)
+p^2_z(\tau)]/2}{\omega_\mathbf{k}(t_0)}d\tau$ in scalar QED. Note that this difference is nonexistent in the QVE.

\begin{figure}[!ht]
\centering
\includegraphics[width=0.45\textwidth]{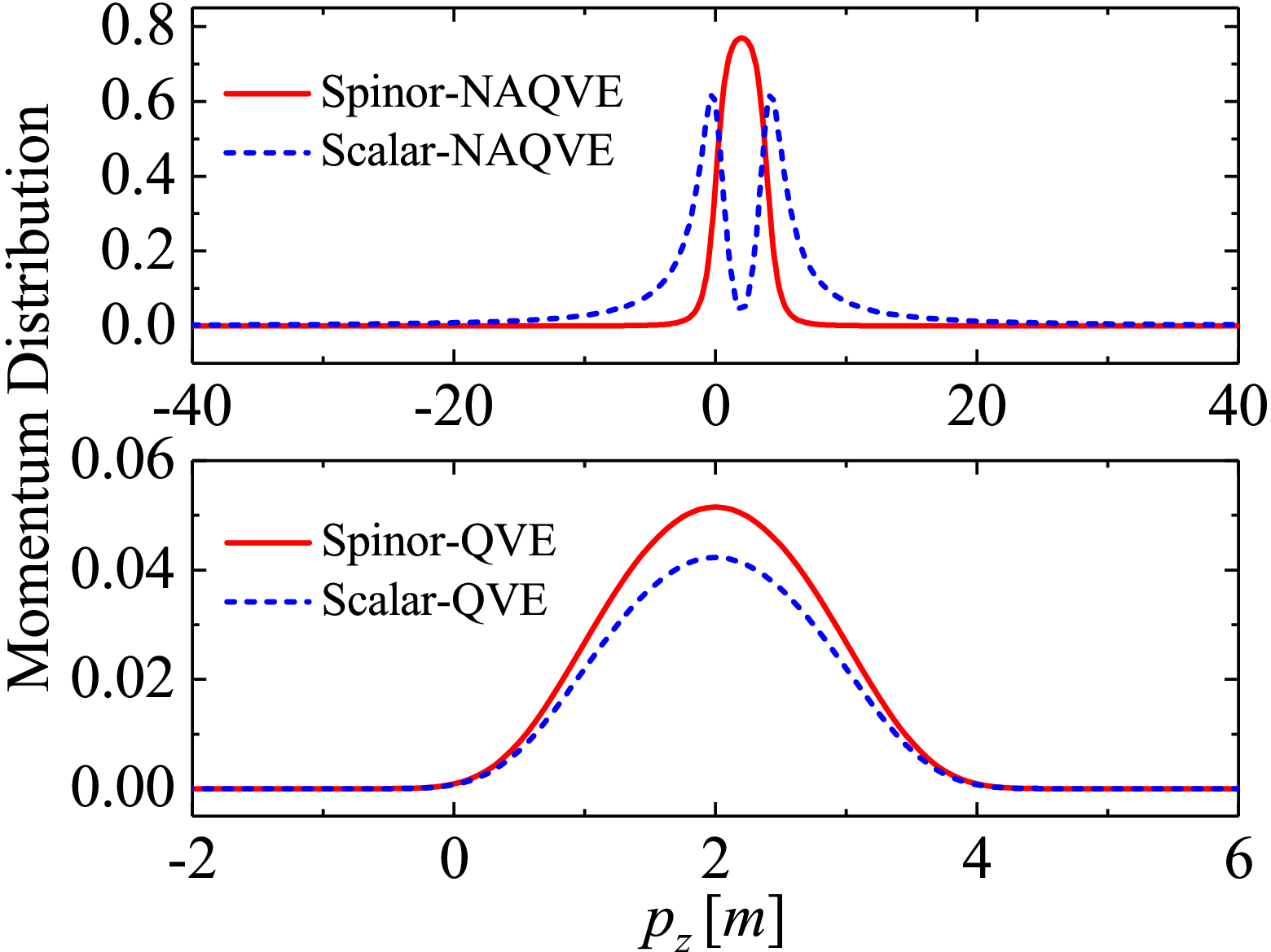}%
\caption{Momentum distributions for NAQVE (upper panel) and QVE (lower panel) in the spinor and scalar QED in the case of $k_x=k_y=0$. The parameters of the field (\ref{eqn:electricfield}) are $E_0=1E_{\rm{cr}}$ and $\tau=2/m$.
\label{fig:Fig5}}
\end{figure}

To clearly see the difference between the NAQVE in spinor and scalar QED, Fig. \ref{fig:Fig4} shows the time evolution of the distribution functions for NAQVE (upper panel) and QVE (lower panel) in the spinor and scalar QED with the momentum $k_x=0$ and $k_y=k_z=0.1m$. It can be seen that, in contrast to the QVE, the distribution functions for the NAQVE in spinor QED and scalar QED are very different. The distribution function in the scalar QED is almost two times that in the spinor QED after the electric field is turned off. This is because the shape of the momentum distribution in these two cases is different, see the upper panel in Fig. \ref{fig:Fig5}. At the position of the maximum distribution function in spinor QED, the distribution function in scalar QED shows a local minimum. In fact, the maximum distribution function in the spinor QED is larger than that in the scalar QED.

In addition, it is found that after the electric field is turned off, the distribution functions in spinor and scalar QED have the same oscillation period but a phase difference of approximately $\pi/2$. This result can be predicted from the last terms on the right-hand side of Eq. (\ref{eqn:Relation1}) and Eq. (\ref{RelationScQED2}). From Eqs. (\ref{eqn:Ck}) and (\ref{CkBoson1}), one can see that, similar to the first factors in the last terms, factors $C^\pm_\mathbf{k}(t)$ are also constants after the electric field is turned off. Therefore, the only reason for the oscillations of distribution functions comes from the exponential factors in the last terms. Then we can find that both of the oscillation periods of distribution functions in spinor and scalar QED are $\pi/\omega_\mathbf{k}(t_f)$, where $t_f$ denotes the time after the electric field is turned off. Moreover, for certain momentum, when the real part of $C^+_\mathbf{k}(t_f)$ plays a leading role in spinor QED, the imaginary part of $C^-_\mathbf{k}(t_f)$ may dominate in scalar QED, and vice versa. So there may be a phase difference of approximately $\pi/2$ for the oscillations of distribution functions in spinor and scalar QED.

In \cite{HebenstreitPRL2009,Hebenstreit2009}, the authors found that the momentum distribution of created pairs presents obvious oscillations for a spatially homogeneous and time-dependent electric field with subcycle structure. Furthermore, at a certain momentum, the momentum distribution shows a local maximum in spinor QED but a local minimum in scalar QED and vice versa, i.e., the out-of-phase behavior. These findings have been explained as the Stokes phenomenon mathematically and a resonance effect in a one-dimensional over-the-barrier quantum mechanical scattering problem physically by an extended WKB approximation \cite{Dumlu2010}. In detail, the oscillatory structure of momentum distribution can be understood as interference effects between multiple pairs of turning points (the solutions of $\omega_\mathbf{k}(t)=0$). In the framework of QVE, for the electric field (\ref{eqn:electricfield}), there is only one pair of turning points dominating pair production. So both in spinor and scalar QED, the momentum distributions have no oscillatory structure; see the lower panel in Fig. \ref{fig:Fig5}. However, for NAQVE, although the momentum distribution in spinor QED has no oscillation, it shows obvious oscillation in scalar QED; see the upper panel in Fig. \ref{fig:Fig5}. Notice that since, for NAQVE, the distribution function still oscillates over time after the electric field is turned off, the momentum distribution is calculated by averaging the distribution function over time. Furthermore, there is an out-of-phase behavior of the momentum distribution between spinor QED and scalar QED. One may think that, for NAQVE, the turning points should be defined as the solutions of $\Omega^\pm_\mathbf{k}(t)=0$, and there may exist more than two pairs of turning points, so the oscillation and out-of-phase behavior are still caused by the interference effect of multiple pairs of turning points. In fact, according to the new definition of turning points, there is still only one pair of turning points, and there should not be any interference effect. This indicates that the extended WKB approximation used in \cite{Dumlu2010} is invalid or should be revised to explain the oscillation and out-of-phase behavior of the momentum distribution for NAQVE.

\begin{figure}[!ht]
\centering
\includegraphics[width=0.45\textwidth]{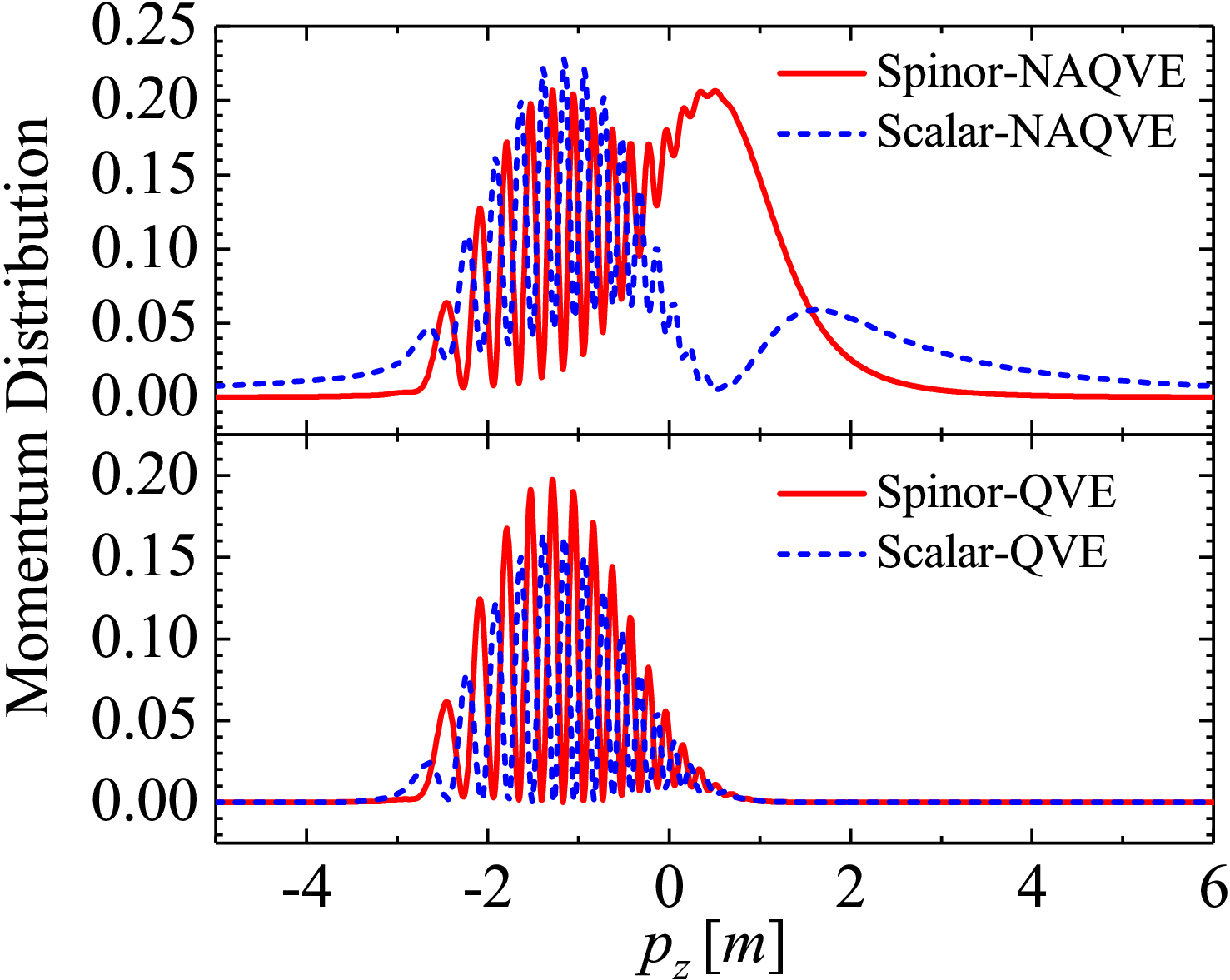}%
\caption{Momentum distributions for NAQVE (upper panel) and QVE (lower panel) in the spinor and scalar QED in the case of $k_x=k_y=0$. The parameters of the field (\ref{eqn:TotalField}) are $E_0=E_{f0}=1E_{\rm{cr}}$, $\tau=2/m$, and $\tau_f=1.5/m$.
\label{fig:Fig6}}
\end{figure}

To see the difference between the oscillatory structures of momentum distributions for NAQVE and QVE, the momentum distribution of created pairs in an asymmetric vector potential consisting of two inequivalent Sauter electric fields is calculated and shown in Fig. \ref{fig:Fig6}. The profile of the electric field is the same as (\ref{eqn:TotalField}), but the field parameters are $E_0=E_{f0}=1E_{\rm{cr}}$, $\tau=2/m$, and $\tau_f=1.5/m$. From the lower panel in Fig. \ref{fig:Fig6}, it can be seen that, for QVE, there are obvious oscillations in the momentum distributions in spinor and scalar QED, and the oscillations present out-of-phase behavior. This result can be well explained by the extended WKB approximation because there are two pairs of turning points for the electric field (\ref{eqn:TotalField}). For NAQVE, within a certain momentum range, the momentum distributions in spinor and scalar QED have similar oscillations as those for QVE. However, beyond this range, the momentum distributions still have out-of-phase behavior that cannot be explained by the extended WKB approximation. Therefore, this type of out-of-phase behavior is different from that caused by the interference effect between multiple pairs of turning points and is worth further study in the future.

\section{SUMMARY AND DISCUSSION}
\label{sec:five}

We have derived the nonadiabatic quantum Vlasov equation in spinor QED by a Bogoliubov transformation and established its relation to the well-known adiabatic QVE by three methods. The first is by an explicitly analytical expression, the second is by the Dirac equation in the V gauge, and the last is by introducing a turn-off electric field. The first two of them give an instantaneous relation while the last one shows an asymptotic relation after turning off the electric field. The establishment of the relation between NAQVE and QVE not only provides us a free choice of the method to study pair production, but also deepens our comprehension of their calculation results. It may also be helpful to exploring the relation between the in-in and in-out formalisms.

By calculating the distribution function varying with time for a specific momentum and comparing it with that in scalar QED, it is found that, similar to the scalar QED, if the frequency of the initial state is not equal to that of the final one the asymptotic distribution function at infinite future also oscillates with time. Furthermore, both of the oscillation periods in spinor and scalar QED are equal to pi divided by the total energy of a particle. The time-averaging momentum distribution after turning off the electric field is also calculated and compared with that in scalar QED. We have found that although there is only one pairs of turning points for the Sauter electric field, the momentum distribution for NAQVE in scalar QED still presents oscillation and the momentum distributions for NAQVE in spinor and scalar QED also show out-of-phase behavior. However, both of the above findings cannot be explained by the Stokes phenomenon, i.e., explained as the interference effect between multiple pairs of turning points. Perhaps it can be explained by revising the WKB approximation. This issue is worthy of further investigation in future. 

Moreover, there are still many signatures studied by QVE worth exploring by NAQVE, such as back reaction, dynamically assisted Schwinger mechanism, pair production in arbitrarily polarized electric fields with or without frequency chirp, node and spiral structures on the momentum spectrum, and so on. These studies will further deepen our understanding of the difference between two versions of QVE and their applications in pair production.


\begin{acknowledgments}
The work is supported by the National Natural Science Foundation of China (NSFC) under Grants No. 11974419 and No. 11705278, and by the Fundamental Research Funds for the Central Universities (No. 2023ZKPYL02).
\end{acknowledgments}

\appendix*
\section{A brief derivation of the nonadiabatic QVE in scalar QED}
\label{app}
For completeness, we give a brief derivation of the nonadiabatic QVE in scalar QED here. The start point is the Klein-Gordon equation for an uniform and time-dependent electric field
\begin{equation}
\big\{\partial_t^2-[\bm{\nabla}-iq\mathbf{A}(t)]^2+m^2\big\}\Phi(\mathbf{x},t)=0.
\end{equation}

The field operator $\hat{\Phi}(\mathbf{x},t)$ can be decomposed as
\begin{equation}\label{FourierDeco1}
\hat{\Phi}(\mathbf{x},t)=\int \frac{d^3k}{(2\pi)^3}\big[\hat{a}_\mathbf{k}\varphi_\mathbf{k}(t)
+\hat{b}^\dagger_{-\mathbf{k}}\varphi_\mathbf{k}^*(t)\big]e^{i\mathbf{k}\cdot\mathbf{x}},
\end{equation}
where $\hat{a}_\mathbf{k}$ and $\hat{b}^\dagger_{-\mathbf{k}}$ are the time-independent annihilation and creation operators which satisfy standard bosonic commutation relations, the complex mode functions $\varphi_\mathbf{k}(t)$ satisfy
\begin{equation}\label{ModeFunction1} \ddot{\varphi}_\mathbf{k}(t)+\omega^2_\mathbf{k}(t)\varphi_\mathbf{k}(t)=0,
\end{equation}
and the Wronskian $\mathrm{Wr}[\varphi^*_\mathbf{k}(t),\varphi_\mathbf{k}(t)]
=\dot{\varphi}^*_\mathbf{k}(t)\varphi_\mathbf{k}(t)
-\varphi^*_\mathbf{k}(t)\dot{\varphi}_\mathbf{k}(t)=-i$, and $\omega^2_\mathbf{k}(t)=[\mathbf{k}-q\mathbf{A}(t)]^2+m^2$.

In the absence of an electric field, the above equation has the plane wave solutions
\begin{equation}\label{PlaneWave}
\varphi_\mathbf{k}^\pm(t)=\frac{1}{\sqrt{2\omega_\mathbf{k}(t_0)}}e^{\mp i\omega_\mathbf{k}(t_0)(t-t_0)},
\end{equation}
where $t_0$ is the initial time and $\omega_\mathbf{k}(t_0)=\{[\mathbf{k}-q\mathbf{A}(t_0)]^2+m^2\}^{1/2}$. These solutions constitute an orthogonal complete set. So the field operator $\hat{\Phi}(\mathbf{x},t)$ can also be expanded as
\begin{equation}\label{FourierDeco2}
\hat{\Phi}(\mathbf{x},t)=\!\!\int \!\frac{d^3k}{(2\pi)^3}\big[\hat{a}_\mathbf{k}(t)\varphi^+_\mathbf{k}(t)
+\hat{b}^\dagger_{-\mathbf{k}}(t)\varphi^-_\mathbf{k}(t)\big]
e^{i\mathbf{k}\cdot\mathbf{x}},
\end{equation}
where $\hat{a}_\mathbf{k}(t)$ and $\hat{b}^\dagger_{-\mathbf{k}}(t)$ are the time-dependent annihilation and creation operators corresponding to the field-free vacuum state and satisfy equal-time commutation relations.

The annihilation and creation operators in Eq. (\ref{FourierDeco1}) and Eq. (\ref{FourierDeco2}) can be associated with each other by a time-dependent Bogoliubov transformation
\begin{eqnarray}\label{Relation1}
\begin{array}{c}
\hat{a}_\mathbf{k}(t)=\alpha_\mathbf{k}(t)\hat{a}_\mathbf{k}
    +\beta^*_\mathbf{k}(t)\hat{b}^\dagger_{-\mathbf{k}} \\
\hat{b}^\dagger_{-\mathbf{k}}(t)=\beta_\mathbf{k}(t)\hat{a}_\mathbf{k}
    +\alpha^*_\mathbf{k}(t)\hat{b}^\dagger_{-\mathbf{k}}
\end{array}
\end{eqnarray}
where $\alpha_\mathbf{k}(t)$ and $\beta_\mathbf{k}(t)$ are the time-dependent Bogoliubov coefficients and satisfy $|\alpha_\mathbf{k}(t)|^2-|\beta_\mathbf{k}(t)|^2=1$ which is the consequence of bosonic commutation relations. Specifically,
\begin{eqnarray}
\alpha_\mathbf{k}(t)&=&i\mathrm{Wr}[(\varphi^+_\mathbf{k}(t))^*,
\varphi_\mathbf{k}(t)]            \nonumber\\
&=&i[\dot{\varphi}_\mathbf{k}(t)-i\omega_\mathbf{k}(t_0)
\varphi_\mathbf{k}(t)]\varphi^-_\mathbf{k}(t),
\label{BogoliubovParametersa}\\
\beta_\mathbf{k}(t)&=&-i\mathrm{Wr}[(\varphi^+_\mathbf{k}(t))^*,
\varphi^*_\mathbf{k}(t)]^* \nonumber\\
&=&-i[\dot{\varphi}_\mathbf{k}(t)+i\omega_\mathbf{k}(t_0)
\varphi_\mathbf{k}(t)]\varphi^+_\mathbf{k}(t). \label{BogoliubovParametersb}
\end{eqnarray}

Using Eqs. (\ref{ModeFunction1}) and (\ref{PlaneWave}), we have
\begin{equation}\label{BogoliubovParameters}
\begin{split}
\dot{\alpha}_\mathbf{k}(t) &=
i[\omega^2_\mathbf{k}(t_0)-\omega^2_\mathbf{k}(t)]
\varphi_\mathbf{k}(t)\varphi^-_\mathbf{k}(t), \\
\dot{\beta}_\mathbf{k}(t) &= -i[\omega^2_\mathbf{k}(t_0)-\omega^2_\mathbf{k}(t)]
\varphi_\mathbf{k}(t)\varphi^+_\mathbf{k}(t).
\end{split}
\end{equation}
According to Eqs. (\ref{BogoliubovParametersa}) and (\ref{BogoliubovParametersb}), the mode functions $\varphi_\mathbf{k}(t)$ can be expressed by the Bogoliubov coefficients as
\begin{equation}\label{Phi0}
\varphi_\mathbf{k}(t)=\varphi^+_\mathbf{k}(t)\alpha_\mathbf{k}(t)
+\varphi^-_\mathbf{k}(t)\beta_\mathbf{k}(t).
\end{equation}
Then Eq. (\ref{BogoliubovParameters}) can be rewritten as
\begin{equation}\label{BogoliubovParametersab}
\begin{split}
\dot{\alpha}_\mathbf{k}(t) &=\frac{i}{2}Q_\mathbf{k}^-(t)[\alpha_\mathbf{k}(t)+\beta_\mathbf{k}(t)
e^{2i\omega_\mathbf{k}(t_0)(t-t_0)}], \\
\dot{\beta}_\mathbf{k}(t) &=-\frac{i}{2}Q_\mathbf{k}^-(t)[\alpha_\mathbf{k}(t)e^{-2i\omega_\mathbf{k}(t_0)(t-t_0)}
+\beta_\mathbf{k}(t)],
\end{split}
\end{equation}
where $Q_\mathbf{k}^-(t)=[\omega^2_\mathbf{k}(t_0)-\omega^2_\mathbf{k}(t)]
/\omega_\mathbf{k}(t_0)$.
Assuming $\widetilde{\alpha}_\mathbf{k}(t)=\alpha_\mathbf{k}(t)\exp[-\frac{i}{2}
\int_{t_0}^{t}Q_\mathbf{k}^-(\tau)d\tau]$ and $\widetilde{\beta}_\mathbf{k}(t)=\beta_\mathbf{k}(t)\exp[\frac{i}{2}
\int_{t_0}^{t}Q_\mathbf{k}^-(\tau)d\tau]$, then Eq. (\ref{BogoliubovParametersab}) becomes
\begin{equation}\label{abtilde}
\begin{split}
\dot{\widetilde{\alpha}}_\mathbf{k}(t) &=\frac{i}{2}Q_\mathbf{k}^-(t)\widetilde{\beta}_\mathbf{k}(t)
e^{2i\theta^-_\mathbf{k}(t_0,t)}, \\
\dot{\widetilde{\beta}}_\mathbf{k}(t) &=-\frac{i}{2}Q_\mathbf{k}^-(t)\widetilde{\alpha}_\mathbf{k}(t)
e^{-2i\theta^-_\mathbf{k}(t_0,t)},
\end{split}
\end{equation}
where $\theta^-_\mathbf{k}(t_0,t)=\int_{t_0}^{t}[\omega^2_\mathbf{k}(t_0)+\omega^2_\mathbf{k}(\tau)
]/[2\omega_\mathbf{k}(t_0)]d\tau$.

The momentum distribution function of particle pairs produced from the initial vacuum by the electric field is defined as
\begin{eqnarray}\label{eqn:NumberDensity}
f^-_{\mathbf{k}}(t)&\equiv&\lim_{V\rightarrow\infty}\frac{\langle \mathrm{vac}|\hat{a}^{\,\dagger}_\mathbf{k}(t)
\hat{a}_\mathbf{k}(t)|\mathrm{vac}\rangle}{V} \nonumber\\
&=&|\beta_\mathbf{k}(t)|^2=|\widetilde{\beta}_\mathbf{k}(t)|^2.
\end{eqnarray}

Similar to the derivation of the nonadiabatic QVE in spinor QED, one can obtain the integro-differential form of the nonadiabatic QVE in scalar QED:
\begin{equation}\label{QVE1}
\begin{split}
\dot{f}^-_\mathbf{k}(t)=\frac{1}{2}Q^-_\mathbf{k}(t)\int_{t_0}^t d t'\, &Q^-_\mathbf{k}(t')[1+2f^-_\mathbf{k}(t')] \\ &\times\cos[2\theta^-_\mathbf{k}(t',t)],
\end{split}
\end{equation}
where $\theta^-_\mathbf{k}(t',t)=\int_{t'}^{t}\Omega^-_\mathbf{k}(\tau)\,d \tau$ and $\Omega^-_\mathbf{k}(\tau)=[\omega^2_\mathbf{k}(t_0)+\omega^2_\mathbf{k}(\tau)
]/[2\omega_\mathbf{k}(t_0)]$. Set
\begin{equation}
g^-_\mathbf{k}(t)=\!\int_{t_0}^t d t'\,Q^-_\mathbf{k}(t')[1+2f^-_\mathbf{k}(t')]\cos[2\theta^-_\mathbf{k}(t',t)]\nonumber
\end{equation}
and
\begin{equation}
h^-_\mathbf{k}(t)=\!\int_{t_0}^t d t'\,Q^-_\mathbf{k}(t')[1+2f^-_\mathbf{k}(t')]\sin[2\theta^-_\mathbf{k}(t',t)],\nonumber
\end{equation}
equation (\ref{QVE1}) can be equivalently transformed into a set of ODEs:
\begin{eqnarray}\label{QVE2}
\dot{f}^-_\mathbf{k}(t)&\!=\!&\frac{1}{2}Q^-_\mathbf{k}(t)g^-_\mathbf{k}(t), \nonumber\\
\dot{g}^-_\mathbf{k}(t)&\!=\!&Q^-_\mathbf{k}(t)[1+2f^-_\mathbf{k}(t)]
-2\,\Omega^-_\mathbf{k}(t)h^-_\mathbf{k}(t)  , \qquad\;\\
\dot{h}^-_\mathbf{k}(t)&\!=\!&2\,\Omega^-_\mathbf{k}(t)g^-_\mathbf{k}(t),\nonumber
\end{eqnarray}
with the initial conditions $f^-_\mathbf{k}(t_0)=g^-_\mathbf{k}(t_0)=h^-_\mathbf{k}(t_0)=0$.

Furthermore, similar to the derivation in Sec. \ref{B}, we can derive the instantaneous relation between the NAQVE and the QVE in scalar QED as
\begin{equation}\label{RelationScQED1}
\begin{split}
f^-_{\mathbf{k}}(t)=&\Big[\frac{\omega^2_\mathbf{k}(t_0)+\omega^2_\mathbf{k}(t)}
{4\omega_\mathbf{k}(t_0)\omega_\mathbf{k}(t)}
+\frac{\dot{\omega}^2_\mathbf{k}(t)}{2\omega_\mathbf{k}(t_0)\omega^2_\mathbf{k}(t)}\Big]
[2F^-_{\mathbf{k}}(t)+1]\\
&-\frac{1}{2}+2\Re\Big\{\Big[\frac{\omega^2_\mathbf{k}(t_0)-\omega^2_\mathbf{k}(t)}
{4\omega_\mathbf{k}(t_0)\omega_\mathbf{k}(t)}
+\frac{\dot{\omega}^2_\mathbf{k}(t)}{2\omega_\mathbf{k}(t_0)\omega^2_\mathbf{k}(t)}\\
&-i\frac{\dot{\omega}_\mathbf{k}(t)}{\sqrt{2\omega_\mathbf{k}(t)}\omega_\mathbf{k}(t_0)}\Big]
C^-_{\mathbf{k}}(t)e^{2i\int_{t_0}^{t}\omega_\mathbf{k}(\tau)d\tau}\Big\},
\end{split}
\end{equation}
where $\dot{\omega}_\mathbf{k}(t)=qE_z(t)p_z(t)/\omega_\mathbf{k}(t)$ is the time derivative of $\omega_\mathbf{k}(t)$, $F^-_{\mathbf{k}}(t)=\lim_{V\rightarrow\infty}\frac{\langle \mathrm{vac}|\hat{\widetilde{a}}^{\,\dagger}_{\mathbf{k}}(t)
\hat{\widetilde{a}}_{\mathbf{k}}(t)|\mathrm{vac}\rangle}{V}$ is the momentum distribution function of created adiabatic bosons, the letters with a tilde denote the creation and annihilation operators of adiabatic bosons,
\begin{equation}\label{CkBoson1}
\begin{split}
C^-_{\mathbf{k}}(t)=&\lim_{V\rightarrow\infty}\frac{\langle \mathrm{vac}|\hat{\widetilde{a}}^{\,\dagger}_{\mathbf{k}}(t)
\hat{\widetilde{b}}^{\,\dagger}_{\mathbf{-k}}(t)
|\mathrm{vac}\rangle}{V}\\
=&\int_{t_0}^tdt'\frac{qE_z(t')p_z(t')}{2\omega^2_\mathbf{k}(t')}
\left[2F^-_\mathbf{k}(t')+1\right]\\
&\hspace{+1cm}\times e^{-2i\int_{t_0}^{t'}
\omega_\mathbf{k}(\tau)d\tau}
\end{split}
\end{equation}
is the time-dependent pair correlation function describing the production of a quasi-boson pair. When the adiabatic condition ($\dot{\omega}_\mathbf{k}(t)/\omega^2_\mathbf{k}(t)\ll1$ and $\ddot{\omega}_\mathbf{k}(t)/\omega^3_\mathbf{k}(t)\ll1$, see \cite{Kluger1998}) is satisfied, especially when the electric field vanishes, Eq. (\ref{RelationScQED1}) can be simplified as
\begin{equation}\label{RelationScQED2}
\begin{split}
f^-_{\mathbf{k}}(t)=&\frac{\omega^2_\mathbf{k}(t_0)+\omega^2_\mathbf{k}(t)}
{4\omega_\mathbf{k}(t_0)\omega_\mathbf{k}(t)}[2F^-_{\mathbf{k}}(t)+1]-\frac{1}{2}\\
&+\frac{\omega^2_\mathbf{k}(t_0)-\omega^2_\mathbf{k}(t)}
{2\omega_\mathbf{k}(t_0)\omega_\mathbf{k}(t)}\Re
\Big\{C^-_{\mathbf{k}}(t)e^{2i\int_{t_0}^{t}\omega_\mathbf{k}(\tau)d\tau}\Big\},
\end{split}
\end{equation}
where
\begin{equation}\label{CkBoson2'}
\begin{split}
C^-_{\mathbf{k}}(t)=\int_{t_0}^tdt'&\frac{qE_z(t')p_z(t')}{2\omega^2_\mathbf{k}(t')}
\left[2F^-_\mathbf{k}(t')+1\right]\\
&\times e^{-2i\int_{t_0}^{t'}\omega_\mathbf{k}(\tau)d\tau}.
\end{split}
\end{equation}
According to the integro-differential form of adiabatic QVE,
\begin{equation}\label{CkBoson2}
\begin{split}
&\Re\Big\{C^-_{\mathbf{k}}(t)
e^{2i\int_{t_0}^{t}\omega_\mathbf{k}(\tau)d\tau}\Big\}\\
=&\int_{t_0}^tdt'\frac{qE_z(t')p_z(t')}{2\omega^2_\mathbf{k}(t')}
\left[2F^-_\mathbf{k}(t')+1\right]\\
&\hspace{+1cm}\times \cos\Big[{2\int_{t'}^{t}
\omega_\mathbf{k}(\tau)d\tau}\Big]\\
=&\frac{\omega^2_\mathbf{k}(t)}{qE_z(t)p_z(t)}\dot{F}^-_\mathbf{k}(t).
\end{split}
\end{equation}
So equation (\ref{RelationScQED2}) can be rewritten as
\begin{equation}\label{RelationScQED2'}
\begin{split}
f^-_{\mathbf{k}}(t)=&\frac{\omega^2_\mathbf{k}(t_0)+\omega^2_\mathbf{k}(t)}
{4\omega_\mathbf{k}(t_0)\omega_\mathbf{k}(t)}[2F^-_{\mathbf{k}}(t)+1]-\frac{1}{2}\\
&+\frac{p_z^2(t_0)-p_z^2(t)}
{2 p_z(t)\dot{p}_z(t)}\frac{\omega_\mathbf{k}(t)}{\omega_\mathbf{k}(t_0)}
\dot{F}^-_\mathbf{k}(t).
\end{split}
\end{equation}


\begin{thebibliography}{99}

\bibitem{Dirac19301}
P. A. M. Dirac, A theory of electrons and protons, Proc. Roy. Soc. Lond. A \textbf{126}, 360 (1930).

\bibitem{Sauter1931}
F. Sauter, \"{U}ber das Verhalten eines Elektrons im homogenen elektrischen Feld nach der relativistischen Theorie Diracs, Z. Phys. \textbf{69}, 742 (1931).

\bibitem{Heisenberg1936}
W. Heisenberg and H. Euler, Folgerungen aus der Diracschen Theorie des Positrons, Z. Phys. \textbf{98}, 714 (1936).

\bibitem{Schwinger1951}
J. Schwinger, On Gauge Invariance and Vacuum Polarization, Phys. Rev. \textbf{82}, 664 (1951).

\bibitem{Xie2017}
B. S. Xie, Z. L. Li, and S. Tang, Electron-positron pair production in ultrastrong laser fields, Matter and Radiation at Extremes \textbf{2}, 225 (2017).

\bibitem{Fedotov2023}
A. Fedotov, A. Ilderton, F. Karbstein, B. King, D. Seipt, H. Taya, and G. Torgrimsson, Advances in QED with intense background fields, Phys. Rep. \textbf{1010}, 1 (2023).


\bibitem{Brezin1970}
E. Br\'{e}zin and C. Itzykson, Pair production in vacuum by an alternating field, Phys. Rev. D \textbf{2}, 1191 (1970).

\bibitem{Hu2010}
H. Y. Hu, C. M\"{u}ller, and C. H. Keitel, Complete QED theory of multiphoton
trident pair production in strong laser fields, Phys. Rev. Lett. \textbf{105}, 080401 (2010).

\bibitem{Ilderton2011}
A. Ilderton, Trident pair production in strong laser pulses, Phys. Rev. Lett. \textbf{106}, 020404 (2011).

\bibitem{Piazza2016}
A. Di Piazza, Nonlinear Breit-Wheeler pair production in a tightly focused laser beam, Phys. Rev. Lett. \textbf{117}, 213201 (2016).

\bibitem{Mackenroth2018}
F. Mackenroth and A. Di Piazza, Nonlinear trident pair production in an arbitrary plane wave: A focus on the properties of the transition amplitude, Phys. Rev. D \textbf{98}, 116002 (2018).


\bibitem{Marinov1977}
M. S. Marinov and V. S. Popov, Electron-positron pair creation from vacuum induced by variable electric field, Fortschr. Phys. \textbf{25}, 373 (1977).

\bibitem{Dumlu2010}
C. K. Dumlu and G. V. Dunne, Stokes Phenomenon and Schwinger Vacuum Pair Production in Time-Dependent Laser Pulses, Phys. Rev. Lett. \textbf{104}, 250402 (2010).

\bibitem{Li2014-1}
Z. L. Li, D. Lu, and B. S. Xie, Multiple-slit interference effect in the time domain for boson pair production, Phys. Rev. D \textbf{89}, 067701 (2014).


\bibitem{Oertel2019}
J. Oertel and R. Sch\"{u}tzhold, WKB approach to pair creation in spacetime-dependent fields: The case of a spacetime-dependent mass, Phys. Rev. D \textbf{99}, 125014 (2019).

\bibitem{Taya2021}
H. Taya, T. Fujimori, T. Misumi, M. Nittaa, and N. Sakai, Exact WKB analysis of the vacuum pair production by time-dependent electric fields, J. High Energ. Phys. \textbf{03}, 82 (2021).

\bibitem{Kohlfurst2022}
C. Kohlf\"{u}rst, N. Ahmadiniaz, J. Oertel, R. Sch\"{u}tzhold, Sauter-Schwinger effect for colliding laser pulses, Phys. Rev. Lett. \textbf{129}, 241801 (2022).


\bibitem{Affleck1982}
I. K. Affleck, O. Alvarez, and N. S. Manton, Pair production at strong coupling in weak external fields, Nucl. Phys. B \textbf{197}, 509 (1982).

\bibitem{Kim2002}
S. P. Kim and D. N. Page, Schwinger pair production via instantons in strong electric fields, Phys. Rev. D \textbf{65}, 105002(2002).

\bibitem{Dunne2005}
G. V. Dunne and C. Schubert, Worldline instantons and pair production in inhomogenous fields, Phys. Rev. D \textbf{72}, 105004 (2005).


\bibitem{Dumlu2011}
C. K. Dumlu and G. V. Dunne, Complex worldline instantons and quantum interference in vacuum pair production, Phys. Rev. D \textbf{84}, 125023 (2011).

\bibitem{Schneider2018}
C. Schneider, G. Torgrimsson, and R. Sch\"{u}tzhold, Discrete worldline instantons, Phys. Rev. D \textbf{98}, 085009 (2018).



\bibitem{Kluger1998}
Y. Kluger, E. Mottola, and J. M. Eisenberg, Quantum Vlasov equation and its Markov limit, Phys. Rev. D \textbf{58}, 125015 (1998).

\bibitem{Schmidt1998}
S. M. Schmidt, D. Blaschke, G. Ropke, S. A. Smolyansky, A. V. Prozorkevich, and V. D. Toneev, A quantum kinetic equation for particle production in the Schwinger mechanism, Int. J. Mod. Phys. E \textbf{7}, 709 (1998).

\bibitem{Alkofer2001}
R. Alkofer, M. B. Hecht, C. D. Roberts, S. M. Schmidt, and D. V. Vinnik, Pair creation and an X-ray free electron laser, Phys. Rev. Lett. \textbf{87}, 193902 (2001).

\bibitem{HebenstreitPRL2009}
F. Hebenstreit, R. Alkofer, G. V. Dunne, and H. Gies, Momentum Signatures for Schwinger Pair Production in Short Laser Pulses with a Subcycle Structure, Phys. Rev. Lett. \textbf{102}, 150404 (2009).

\bibitem{Hebenstreit2009}
F. Hebenstreit, R. Alkofer, G. V. Dunne, and H. Gies, Quantum statistics effect in Schwinger pair production in short laser pulses, arXiv:0910.4457 [hep-ph].

\bibitem{Kohlfurst2014}
C. Kohlf\"{u}rst, H. Gies, and R. Alkofer, Effective mass signatures in multiphoton pair production, Phys. Rev. Lett. \textbf{112}, 050402 (2014).



\bibitem{Bialynicki-Birula1991}
I. Bialynicki-Birula, P. G\'{o}rnicki, and J. Rafelski, Phase-space structure of the Dirac vacuum, Phys. Rev. D \textbf{44}, 1825 (1991).

\bibitem{Hebenstreit2010}
F. Hebenstreit, R. Alkofer, and H. Gies, Schwinger pair production in space- and time-dependent electric fields: Relating the Wigner formalism to quantum kinetic theory, Phys. Rev. D \textbf{82}, 105026 (2010).

\bibitem{Hebenstreit2011}
F. Hebenstreit, R. Alkofer, and H. Gies, Particle self-bunching in the Schwinger effect in spacetime-dependent electric fields, Phys. Rev. Lett. \textbf{107}, 180403 (2011).

\bibitem{Blinne2014}
A. Blinne and H. Gies, Pair production in rotating electric fields, Phys. Rev. D \textbf{89}, 085001 (2014).

\bibitem{Li2017}
Z. L. Li, Y. J. Li, and B. S. Xie, Momentum Vortices on Pairs Production by Two Counter-Rotating Fields, Phys. Rev. D \textbf{96}, 076010 (2017).


\bibitem{Kohlfurst2020}
C. Kohlf\"{u}rst, Effect of time-dependent inhomogeneous magnetic fields on the particle momentum spectrum in electron-positron pair production, Phys. Rev. D \textbf{101}, 096003 (2020).

\bibitem{Cheng2010}
T. Cheng, Q. Su, and R. Grobe, Introductory review on quantum field theory with space –time resolution, Contemp. Phys. \textbf{51}, 315 (2010).

\bibitem{Jiang2012}
M. Jiang, W. Su, Z. Q. Lv, X. Lu, Y. J. Li, R. Grobe, and Q. Su, Pair creation enhancement due to combined external fields, Phys. Rev. A \textbf{85}, 033408 (2012).

\bibitem{Su2012}
Q. Su, W. Su, Q. Z. Lv, M. Jiang, X. Lu, Z. M. Sheng, and R. Grobe, Magnetic control of the pair creation in spatially localized supercritical fields, Phys. Rev. Lett. \textbf{109}, 253202 (2012).

\bibitem{Dong2017}
S. S. Dong, M. Chen, Q. Su, and R. Grobe, Optimization of spatially localized electric fields for electron-positron pair creation, Phys. Rev. A \textbf{96}, 032120 (2017).


\bibitem{Su2019}
Q. Su and R. Grobe, Dirac vacuum as a transport medium for tnformation, Phys. Rev. Lett. \textbf{122}, 023603 (2019).

\bibitem{Schmidt1999}
S. Schmidt, D. Blaschke, G. R\"{o}pke, A. V. Prozorkevich, S. A. Smolyansky, and V. D. Toneev, Non-Markovian effects in strong-field pair creation, Phys. Rev. D \textbf{59}, 094005 (1999).

\bibitem{Bloch1999}
J. C. R. Bloch, V. A. Mizerny, A. V. Prozorkevich, C. D. Roberts, S. M. Schmidt, S. A. Smolyansky, and D. V. Vinnik, Pair creation: Back reactions and damping, Phys. Rev. D \textbf{60}, 116011 (1999).

\bibitem{Jiang2023}
R. Z. Jiang, C. Gong, Z. L. Li, and Y. J. Li, Backreaction effect and plasma oscillation in pair production for rapidly oscillating electric fields, Phys. Rev. D \textbf{108}, 076015 (2023).

\bibitem{Orthaber2011}
M. Orthaber, F. Hebenstreit, and R. Alkofer, Momentum spectra for dynamically assisted Schwinger pair production, Phys. Lett. B \textbf{698}, 80 (2011).

\bibitem{Nuriman2012}
A. Nuriman, B. S. Xie, Z. L. Li, and D. Sayipjamal, Enhanced electron–positron pair creation by dynamically assisted combinational fields, Phys. Lett. B \textbf{717}, 465 (2012).

\bibitem{Kohlfurst2013}
C. Kohlf\"{u}rst, M. Mitter, G. von Winckel, F. Hebenstreit, and R. Alkofer, Optimizing the pulse shape for Schwinger pair production, Phys. Rev. D \textbf{88}, 045028 (2013).

\bibitem{Hebenstreit2014}
F. Hebenstreit and F. Fillion-Gourdeau, Optimization of Schwinger pair production in colliding laser pulses,  Phys. Lett. B \textbf{739}, 189 (2014).

\bibitem{Li2014}
Z. L. Li, D. Lu, B. S. Xie, L. B. Fu, J. Liu, and B. F. Shen, Enhanced pair production in strong fields by multiple-slit interference effect with dynamically assisted Schwinger mechanism, Phys. Rev. D \textbf{89}, 093011 (2014).

\bibitem{Dumlu20101}
C. K. Dumlu, Schwinger vacuum pair production in chirped laser pulses, Phys. Rev. \textbf{82}, 045007 (2010).

\bibitem{Jiang2013}
M. Jiang, B. S. Xie, H. B. Sang, and Z. L. Li, Enhanced electron-positron pair creation by the frequency chirped laser pulse, Chin. Phys. B \textbf{22}, 100307 (2013).

\bibitem{Abdukerim2017}
N. Abdukerim, Z. L. Li, and B. S. Xie, Enhanced electron-positron pair production by frequency chirping in one- and two-color laser pulse fields, Chin. Phys. B \textbf{26}, 020301 (2017).

\bibitem{GongPRD2020}
C. Gong, Z. L. Li, B. S. Xie, and Y. J. Li, Electron-positron pair production in frequency modulated laser fields, Phys. Rev. D \textbf{101}, 016008 (2020).


\bibitem{Dumlu2009}
C. K. Dumlu, Quantum kinetic approach and the scattering approach to vacuum pair production, Phys. Rev. D \textbf{79}, 065027 (2009).

\bibitem{Li2019}
Z. L. Li, B. S. Xie, and Y. J. Li, Boson pair production in arbitrarily polarized electric fields, Phys. Rev. D \textbf{100}, 076018 (2019).

\bibitem{Li2021}
Z. L. Li, C. Gong, and Y. J. Li, Study of pair production in inhomogeneous two-color electric fields using the computational quantum field theory, Phys. Rev. D \textbf{103}, 116018 (2021).

\bibitem{Kim2011}
S. P. Kim and C. Schubert, Nonadiabatic quantum Vlasov equation for Schwinger pair production, Phys. Rev. D \textbf{84}, 125028 (2011).

\bibitem{Huet2014}
A. Huet, S. P. Kim, and C. Schubert, Vlasov equation for Schwinger pair production in a time-dependent electric field, Phys. Rev. D \textbf{90}, 125033 (2014).

\bibitem{Campos1994}
A. Campos and E. Verdaguer, Semiclassical equations for weakly inhomogeneous cosmologies, Phys. Rev. D \textbf{49}, 1861 (1994).

\bibitem{Higuchi2011}
A. Higuchi, D. Marolf, and I. A. Morrison, Equivalence between Euclidean and in-in formalisms in de Sitter QFT, Phys. Rev. D \textbf{83}, 084029 (2011).

\bibitem{Ota2023}
A. Ota, M. Sasaki, and Y. Wang, One-loop tensor power spectrum from an excited scalar field during inflation, Phys. Rev. D \textbf{108}, 043542 (2023).

\bibitem{Braun1999}
J. W. Braun, Q. Su, and R. Grobe, Numerical approach to solve the time-dependent Dirac equation, Phys. Rev. A \textbf{59}, 604
(1999).

\bibitem{Mocken20041}
G. R. Mocken and C. H. Keitel, Quantum dynamics of relativistic electrons, J. Comput. Phys. \textbf{199}, 558 (2004);

\bibitem{Mocken20042}
G. R. Mocken and C. H. Keitel, FFT-split-operator code for solving the Dirac equation in 2+1 dimensions, Comput. Phys. Commun. \textbf{178}, 868 (2008).





\end{thebibliography}
\end{document}